\documentclass[twocolumn,pre,showpacs,amsfonts,amsmath,assymb,eufrak
,longbibliography]{revtex4-1}
\usepackage{epsfig}
\usepackage{color}
\usepackage{bm}
\usepackage[
colorlinks=true, 
pdfstartview=FitV, 
linkcolor=blue, 
citecolor=magenta, 
urlcolor=blue, 
bookmarks=true,
bookmarksnumbered=true,
pdftitle={},
pdfauthor={}
]{hyperref}
\usepackage{epsfig,amsopn}
\usepackage{graphicx}

\usepackage{amssymb}
\usepackage{amsmath}
\usepackage{color}
\usepackage{subfigure}
\usepackage{textcomp}
\usepackage[utf8]{inputenc}
\definecolor{pink}{rgb}{1,0,0.6}

\newcommand\ra{\rangle}
\newcommand\la{\langle}
\newcommand\nn{\nonumber}
\newcommand\f{\frac}
\newcommand\p{\partial}
\usepackage{grffile}
\graphicspath{{./Figures/}{./Hydro/}{./SpaceAverage/}{./FiguresCan/}}
\usepackage{textcomp}
\usepackage{amsmath}

\DeclareMathOperator\erf{erf}

\newcommand{\be}{\begin{equation}}
\newcommand{\ee}{\end{equation}}
\newcommand{\bea}{\begin{eqnarray}}
\newcommand{\eea}{\end{eqnarray}}

\begin{document}

\newcommand{\titlename}{The Taylor-von Neumann-Sedov blast-wave solution: comparisons with microscopic simulations of a one-dimensional gas}

\title{\titlename}% Force line breaks with \\

\author{ Santhosh Ganapa$^1$, Subhadip Chakraborti$^1$, P. L. Krapivsky$^2$ and Abhishek Dhar$^1$}

\affiliation{$^1$International Centre for Theoretical Sciences, Tata Institute of Fundamental Research, Bengaluru 560089, India}%
\affiliation{$^2$Department of Physics, Boston University, Boston, MA 02215, USA}

\date{\today}
\begin{abstract} 
We study the response of an infinite system of point particles on the line initially at rest on the instantaneous release of energy in a localized region. 
We make a detailed comparison of the hydrodynamic variables predicted by Euler equations for non-dissipative ideal compressible gas and the results of direct microscopic simulations.  At long times the profiles of the three conserved variables evolve to  self-similar scaling forms, with a scaling exponent as predicted by the Taylor-von Neumann-Sedov (TvNS) blast-wave solution. The scaling functions obtained from the microscopic dynamics show a remarkable agreement with the TvNS predictions, except at the blast core, where the TvNS solution predicts a diverging temperature which is not observed in simulations. We show that the effect of heat conduction becomes important and present results from a numerical solution of the full Navier-Stokes-Fourier equations. A different scaling form is observed in the blast core and this is carefully analyzed. Our microscopic model is the one-dimensional alternate mass hard-particle gas which has the ideal gas equation of state but is non-integrable and known to display fast equilibration.
\keywords{Hard particle gas, Hydrodynamics, Nonequilibrium dynamics}
\end{abstract}
%\tableofcontents
\maketitle

 \section{Introduction}
\label{sec:Intro}
The evolution of a blast wave emanating from an intense explosion has drawn much interest since it was first studied to understand the mechanical effect of bombs. The rapid release of large amount of energy in a localized region produces a surface of discontinuity beyond which the quantities concerned like density, velocity and temperature fields change discontinuously \cite{LandauBook,ZeldovichBook}. The evolution of this surface of discontinuity, also known as the shock front, has been studied extensively in the last few decades. A solution of this 
problem was first proposed by Taylor \cite{Taylor19501,Taylor19502}, von Neumann \cite{VonNeumann1963} and Sedov \cite{Sedov1946,Sedov2014} and referred to as the Taylor-von Neumann-Sedov  (TvNS) solution. These solutions have a self-similar form at large times ($t$). Consider the case where initially the fluid has a density $\rho_\infty$ and is at zero temperature and  we inject a large amount of energy $E $ into a localized region.  In arbitrary dimensions, $d$, assuming a radially symmetric explosion, with $r$ as the distance from the centre of the explosion, it can be shown that the hydrodynamic equations for the density field, $\rho(r,t)$, velocity field, $v(r,t)$, and the temperature field, $T(r,t)$,  admit a  scaling solution at large times of the form: 
\begin{subequations}
\begin{align}
\rho(r,t)&=\rho_\infty G[{r}/{R(t)}],\label{TvNS1} \\
v(r,t)&=\f{r}{t} V[{r}/{R(t)}],\label{TvNS2}\\
T(r,t)&=  \mu \f{r^2}{t^2} Z [{r}/{R(t)}], \label{TvNS3}
\end{align}
\end{subequations}
for $r< R(t)$ and where $R(t)$ is the position of the shock front, $\mu$ is a mass scale.  Using the condition of conservation of energy, it can then be shown~\cite{BarenblattBook} that the radius of the shock front, $R(t)$, scales with time $t$ in $d$ dimensions as
\be
\label{shock_front}
R(t) =  \left(\f{E t^2}{A \rho_\infty}\right)^\frac{1}{d+2},
\ee
where $A$ is a dimensionless constant factor. This scaling law for $R(t)$ has been verified in several experiments such as nuclear explosions \cite{Taylor19501,Taylor19502}, astrophysical blast waves \cite{McKee1988},  laser-driven blast waves in gas jets \cite{Ditmire2001}, plasma \cite{Smith2005}, and recently in granular materials \cite{Kellay2009,Kellay2013}. 

The TvNS analysis is based on the solution of the hydrodynamic Euler equations and an interesting and important question is regarding the comparison of these results with those  from microscopic simulations of the underlying many-particle system following Hamiltonian dynamics. Somewhat surprisingly, such a detailed comparison has been attempted only quite recently.  One of the first studies  explored the kinetics of a blast in a homogeneous fluid of elastic hard spheres (for $d=1$ and $d=2$)~\cite{Antal2008} and verified the TvNS scaling in Eq.~\eqref{shock_front}. More detailed studies to  observe the scaling functions, defined through Eqs.~(\ref{TvNS1}-\ref{shock_front}), in molecular dynamic simulations have recently been performed for hard sphere systems in various dimensions~\cite{Jabeen2010,Barbier2015,Barbier2015a,Barbier2016,Joy2017,Joy2018,Joy2019}, both with elastic and inelastic collisional dynamics.  For the case where energy is conserved, the study in \cite{Barbier2016} for a two-dimensional system found reasonable agreement between the TvNS solution and microscopic simulations when  the ambient density is not too large. On the other hand, more extensive simulations of hard sphere systems in three dimensions~\cite{Joy2018} and in two dimensions~\cite{Joy2019} found significant departures of the simulation results from the  TvNS solution, both near the  centre and at the shock front. While the scaling form seems robust, the scaling functions $G,V,Z$ appear to be different from the TvNS prediction. These studies used equations of states that included upto $10$th order virial coefficients. It was suggested that  possible reasons for the differences could be the lack of local equilibration and the contribution of dissipative terms (viscosity and heat conduction) in the hydrodynamic equations  (that are not included  in the TvNS analysis). Thus the agreement between molecular dynamics simulations of a Hamiltonian system and  hydrodynamics remains an open question.

 In the present paper, we address this question by studying the one-dimensional version of the hard-sphere system. We consider particles with hard point interactions so that the equation of state takes the simple ideal gas form. This example  was already discussed in \cite{Antal2008,Barbier2015a} who noted that one needs to take a system where the masses of alternate particles take different values, say $m_1$ and $m_2$. For the equal mass case the system is integrable and one does not expect the standard hydrodynamic description to work. For the alternate mass 
hard particle (AHP) gas, the scaling $R(t) \sim t^{2/3}$ has been  verified~\cite{Antal2008,Barbier2015a}, however, there are no results for  the scaling functions.  Independently, the AHP gas has been investigated very extensively in the context of the breakdown of Fourier's law of heat conduction in one dimension~\cite{Garrido2001,Dhar2001,Grassberger2002,Casati2003,Cipriani2005,Chen2014,Hurtado2016,lepri2020}. 
A remarkable finding there is that the thermal conductivity $\kappa$ diverges with system size $L$ as $\kappa \sim L^{1/3}$. Some theoretical understanding of this has been achieved using nonlinear fluctuating hydrodynamics~\cite{Narayan2002,Van2012,Mendl2013,Spohn2014}, which makes very detailed predictions for the form of equilibrium dynamical correlation functions. Two microscopic models where these predictions have been verified are the AHP gas~\cite{Mendl2013,Mendl2014} and the Fermi-Pasta-Ulam-Tsingou chain~\cite{Das2014}. As far as the evolution of macroscopic profiles is considered, the AHP gas was used recently~\cite{Mendl2017} to study the so-called Riemann problem~\cite{RiemannBook} where one considers an infinite system with step initial conditions for the three conserved fields. A comparison was made between the simulations and the predictions from the Euler equations. Note that the blast-wave initial conditions lead to a different class of self-similar solutions than those that one gets for the Riemann problem [ballistic scaling with $R(t) \sim t$]. Finally the hard-particle gas has been investigated earlier in the context of breakdown of the hydrodynamic description in one dimension~\cite{Kadanoff1995,Hurtado2006}. In \cite{Hurtado2006} it was noted that the field profiles in front of a rapidly moving piston was not in accordance to hydrodynamic predictions.

Here we provide a detailed study of the evolution of the blast wave initial condition in a one-dimensional fluid with an ideal gas equation of state. We perform extensive molecular dynamics simulations of the AHP gas to compute the evolution of the profiles of density, velocity and temperature fields and thereby extract the scaling forms obtained in the long time limit. We make comparisons with the scaling solution predicted from the TvNS solution which can be numerically obtained with very high precision. One of our main finding is that the simulation results agree remarkably well with the predictions of the TvNS theory except for some  disagreements near the blast core. A different scaling form of the hydrodynamics fields is observed at the core and this is understood as arising from dissipative terms (viscosity and heat conduction) in the hydrodynamic equations which are not present in the Euler description. We then present a detailed numerical and analytical study of the full Navier-Stokes-Fourier (NSF) equations to understand the core region. Our key finding is that the size of the hot core, $X(t)$, grows algebraically with an unusual exponent (smaller of course than the exponent $2/3$ characterizing the position $R(t)$ of the shock wave): 
\begin{equation}
X\sim t^\frac{38}{93}\,,    \qquad    R\sim t^\frac{2}{3}.                        \label{coresize}              
\end{equation}
In contrast to the TvNS solution which predicts a divegence of the temperature in the core, which is quite unphysical, the dissipative equations lead to a finite temperature. For the temperature and density at the core centre we find
\begin{equation}
T_0 \sim t^{-\frac{50}{93}}\,, \qquad \rho_0 \sim t^{-\frac{4}{31}}.  
\label{corescaling}
\end{equation}
We notice that in addition to curing the chief problem of the TvNS solution — the divergence of the temperature — it also cures a lesser problem of the TvNS solution, viz.  the vanishing of the density at the center of explosion.

The paper is organized as follows. In Sec.~\ref{sec:model} we describe the model, the definition of the various conserved densities and the specification of initial conditions. In Sec.~\ref{sec:tvnsscaling}, we discuss the TvNS scaling solution of the Euler equations for an ideal fluid in arbitrary dimensions, emphasing in particular the case for the monoatomic gas in one dimension, which is the focus of this study. In Sec.~\ref{sec:micro}, we show the results of microscopic simulations for the AHP gas and a comparison with the TvNS solution. In  Sec.~\ref{sec:hydro} we study the blast core region through an analysis of the Navier-Stokes-Fourier equations.   We provide a discussion of the main results and remaining open questions in Sec.~\ref{sec:conclusion}.

\section{Microscopic model, definitions of the macroscopic fields and specification of initial conditions}
\label{sec:model}
We study a one-dimensional  gas of $N$ hard point particles moving inside a box  $(-L/2,L/2)$. The only interactions between particles is through  point collisions between nearest neighbors which conserve  energy and momentum and also the ordering of the particles. In between collisions the particles move ballistically with constant speeds. Let $q_j$, $p_j$, and $m_j$ be respectively the position, momentum, and mass of the $j^{th}$ particle with $j=1,2,\ldots,N$. The particles are ordered $-L/2 \leq q_1 \leq q_2,\ldots q_N \leq  L/2$. 
The Hamiltonian  is given by
%%%%%%%%%%%%%%%%%%%%%%%%%%%%%%%%%%%%%%%%%%
\be
{\cal H} = \sum_{j=1}^N \frac{p_j^2}{2m_j} + \sum_{j=1}^{N+1} V(q_j-q_{j-1}),
\ee
%%%%%%%%%%%%%%%%%%%%%%%%%%%%%%%%%%%%%%%%%%
where $q_0=-L/2, q_{N+1}=L/2$ denote the fixed walls and  the interaction potential,  $V(q)$, between two particles is taken to be
%%%%%%%%%%%%%%%%%%%%%%%%%%%%%%%%%%%%%%%%%%
\begin{eqnarray}
V(q) =
\left\{
\begin{array}{ll}
\infty    & {\rm for}~ q \leq 0 \\
0         & {\rm for}~ q > 0.
\end{array}
\right.
\end{eqnarray}
%%%%%%%%%%%%%%%%%%%%%%%%%%%%%%%%%%%%%%%%%%
We consider here the collision between two particles to be perfectly elastic and instantaneous. If a particle with mass $m_1$ and velocity $v_1$ collides with another particle with mass $m_2$ and velocity $v_2$ then after the collision, the velocities of the first and the second particles respectively are given by
%%%%%%%%%%%%%%%%%%%%%%%%%%%%%%%%%%%%%%%%%%
\bea 
\label{elasticcollision1}
v_1^\prime = \frac{2m_2 v_2 + v_1(m_1-m_2)}{m_1+m_2}, \\
v_2^\prime = \frac{2m_1 v_1 + v_2(m_2-m_1)}{m_1+m_2}. \label{elasticcollision2}
\eea
%%%%%%%%%%%%%%%%%%%%%%%%%%%%%%%%%%%%%%%%%%
These can be derived from the conservation of momentum and energy. The particles at the ends on colliding with the fixed walls are simply reflected.

If all particles have equal masses, one  notices from the above equations that after the collision, $v_1^\prime = v_2, v_2^\prime = v_1 $, i.e.  
the particles merely exchange velocities.  This means that the system effectively behaves as a non-interacting system. In fact there are $N$ conserved quantities in the system, which for a periodic ring are the quantities $I_s=\sum_{i=1}^N p_i^s$, for $s=1,2,\ldots,N$.  Hence this system is integrable and we can explicitly write the solution in closed form. Taking the masses to be unequal typically breaks integrability. A well-studied example is the  alternate mass system \cite{Dhar2001,Grassberger2002,Casati2003,Cipriani2005,Antal2008,Chen2014,Hurtado2016,lepri2020} where we choose all odd numbered particles to have mass $m_1$ and all even particles $m_2$.  In this case the only known conserved quantities are the total number of particles, the total momentum and the total energy of the system and we expect a hydrodynamic description in terms of the corresponding conserved fields namely, the mass density field $\rho(x,t)$, the momentum density field $p(x,t)$ and energy density field $E(x,t)$. 

The  hydrodynamic fields, $(\rho,p,E)$ (and the related derived fields $v,e$), are related to the microscopic variables  as
\begin{align}
\rho(x,t)&=\left\langle\sum_{j=1}^N  m_j \delta(q_j(t) - x)\right\rangle, \\
p(x,t)&=\rho(x,t)v(x,t)=\left\langle\sum_{j=1}^N  m_j v_j \delta(q_j(t) - x)\right\rangle, \\
E(x,t)&=\rho(x,t)e(x,t)=\left\langle\sum_{j=1}^N  \frac{1}{2} m_j v_j^2 \delta(q_j(t) - x)\right\rangle.
\end{align}
In the above equations, $\la ...\ra$ indicates an average over an initial distribution of microstates that correspond to the same initial macrostate. Alternatively one can consider the empirical  fields constructed from a single microscopic evolution by performing a spatial averaging process. We divide space into boxes of size $\ell$ which are small compared to the full system size but still typically contain a large number of particles. The empirical  fields are defined as
\begin{align}
&\bar{\rho}(x,t)=\f{1}{\ell}\sum_{j=1}^N m_j \delta[q_j(t),x], \label{br}\\
&\bar{p}(x,t)=\bar{\rho}(x,t)\bar{v}(x,t)=\f{1}{\ell}\sum_{j=1}^N  m_j v_j \delta[q_j(t), x], \label{bp} \\
&\bar{E}(x,t)=\bar{\rho}(x,t)\bar{e}(x,t)=\f{1}{\ell}\sum_{j=1}^N \frac{1}{2} m_j v_j^2 \delta[q_j(t),x], \label{bE}
\end{align}
where $\delta[q_j,x]$ is an indicator function taking value $1$ if $ x-\ell/2 \leq q_j \leq x+\ell/2$ and zero otherwise. In the limit of large enough $\ell$ it  is expected, from typicality arguments, that the empirical  fields  and the ensemble averaged fields should be identical and we will verify this in our study.   

For a non-integrable system, it is expected that the evolving system is in local thermal equilibrium and the three fields should contain the local thermodynamic information at any space-time point $x,t$. To connect to thermodynamics we define other local fields given in terms of the basic fields. First, the internal energy per unit mass is given by
\begin{align}
\epsilon(x,t)= e-\f{v^2}{2}. \label{therm1}
\end{align}
Our system is described by ideal gas thermodynamics hence, the pressure and temperature are given as follows:
\begin{align}
T(x,t)&= 2 \mu \epsilon(x,t),\nn  \\
P(x,t)&= 2 \rho(x,t) \epsilon(x,t)=  \frac{1}{\mu} \rho(x,t) T(x,t),  \label{therm2}
\end{align}
where we have set Boltzmann's constant $k_B=1$.

{\bf  ``Blast-wave'' initial conditions}: We consider an initial macrostate where the gas has a  finite uniform density $\rho_\infty$, zero flow velocity $v$, and is at zero temperature everywhere except in a region of width $\sigma$ centred at $x=0$. This is the region of the blast and has a specified finite temperature profile. In our numerical simulations we considered two different profiles for the initial energy:
\begin{align}
&{\rm (i)~Gaussian~profile}~~E(x,0) = \frac{E }{\sqrt{2\pi\sigma^2}}e^{-x^2/{2\sigma^2}}, \\
&{\rm (ii)~Box~profile}~~E(x,0)=
\begin{cases}\f{E }{2 \sigma},~-\sigma < x < \sigma,\\
0,~~|x| > \sigma.    
\end{cases}
~\label{eq:ezero}
\end{align} 
In both cases the total energy of the blast is $E $. We now describe how one can  realize these macrostates in the microscopic simulations of the AHP gas. Different choices of the initial ensemble of microstates can lead to the same average macroscopic profile. Here we discuss two possible ensemble choices.

\noindent  {\bf (A) Ensemble where  energy and momentum are fixed on average}: In this case the two initial macrostates can be realized as follows:

(i)  Gaussian temperature profile ---  First distribute 
 $N$ ordered  particles numbered say as $i=-N/2+1,-N/2+2,\ldots,N/2$ uniformly between $x = -L/2$ to $L/2$,  so that the number density is $n_0=N/L=\rho_\infty/\mu$, where $\mu=(m_1+m_2)/2$ is the mean mass and $\rho_\infty$ the ambient mass density of the gas.  
For the $N_c$ centre particles with $i=-N_c/2+1, -N_c/2+2,\ldots N_c/2$, we choose their velocities from the Maxwell distribution, $Prob(v_i)= \sqrt{m_i/(2 \pi T)}  e^{-m_i v_i^2/(2 T)}$, where $T=2 E /N_c$. Set the velocities of all other particles to zero. The size of the initial blast is thus approximately $s=N_c/n_0$.

We note that $E(x,0)= \la \sum_{i=1}^N  m_i v_i(0)^2/2 \delta(x-q_i(0)) \ra $.
 Since, for large $N$,  each particle's position is a Gaussian with mean $\bar{q}_i=i/n_0$ and variance $\sigma^2=L/(4 n_0)$, we then get
\begin{align}
E(x,0) &= \sum_{i=-N_c/2+1}^{N_c/2} \f{T}{ 2 \sqrt{2 \pi \sigma^2}}  e^{-(x-\bar{q}_i)^2/ (2 \sigma^2)} \nn \\
&\approx \f{n_0 T}{2} \int_{-s/2}^{s/2} dy   \f{e^{-(x-y)^2/ (2 \sigma^2)}}{  \sqrt{2 \pi \sigma^2}} \nn \\
&=\f{n_0 T}{4} \left[ \erf{\left(\f{s-2x}{2 \sqrt{2} \sigma}\right)} +  \erf{\left(\f{s+2x}{2 \sqrt{2} \sigma}\right)}  \right]. \label{Eerf}  
\end{align}
When $s\ll \sigma$, Eq.~\eqref{Eerf} simplifies to  
\begin{align}
E(x,0)\approx E  \f{e^{-x^2/ (2 \sigma^2)}}{  \sqrt{2 \pi \sigma^2}}. \label{Egauss}
\end{align}

\noindent (ii) { Box temperature profile} --- In this case  we first distribute $(N-N_c)/2$ particles uniformly between  $(-L/2,-\sigma/2)$ and the remaining $(N-N_c)/2$ particles  between $(\sigma/2, L/2)$. The remaining $N_c$ particles are distributed uniformly in the interval $(-\sigma/2,\sigma/2)$.  The mean number density is again $n_0=N/L$. As before  we choose the velocities of the $N_c$ particles in the centred region from the Maxwell distribution, $Prob(v_i)= \sqrt{m_i/(2 \pi T)}  e^{-m_i v_i^2/(2 T)}$, where $T=2 E /N_c$, while  the velocities of all other particles is set to zero.

{\bf (B) Ensemble where  initial energy is fixed exactly to $E $ and momentum exactly to $0$}: In this case we follow the same protocol as before of distributing $N$ particles in space. However we now  choose the velocities of only $N_c/2$ particles ($-N_c/2+1,-N_c/2+2,\ldots,0$), in the centred region from the Maxwell distribution, $Prob(v_i)= \sqrt{m_i/(2 \pi T)}  e^{-m_i v_i^2/(2 T)}$, where $T=2 E /N_c$. For every realization, we rescale these velocities by a constant factor such that the total energy of the $N_c/2$ particles is exactly $E /2$. Each of the remaining particles ($i=1,2,\ldots,N_c/2$) are assigned momentum as $p_i= p_{-i+1}$. Thus we have an ensemble of initial microstates which all have total energy exactly equal to $E $ and total momentum is exactly zero.

{\bf Simulation details}: 
In  all our simulations we took $m_1=1,\ m_2=2$  (so that $\mu=(m_1+m_2)/2=1.5$), $\rho_\infty=1.5$, $E =32$ and $N_c=32$. In our  largest simulations we took $N=24000$, $L=24000$ and averaged over  an ensemble of ${\cal{R}}=10^5$ initial conditions.  For each microscopic initial condition, we evolved the system with the Hamiltonian dynamics. The molecular dynamics simulations for this  can be done efficiently using  an event-driven algorithm which updates the system between successive collisions.

\section{TvNS scaling equations and their solution}
\label{sec:tvnsscaling}
In this section we present the exact solution of the Euler equations for the ideal gas with blast wave initial conditions. 
The blast, caused by the instantaneous release of energy $E$ in a very small region, creates a spherical shock wave propagating through the quiescent gas. The blast is infinitely strong if the pressure behind the shock wave can be neglected. In normal conditions, this is valid up to a certain time; for the gas at zero temperature, it remains valid forever. Below we consider the infinitely strong blast. We begin by discussing the general case for a blast wave in $d$ spatial dimensions for which, because of  the radial symmetry of the problem, we get  three Euler equations for the hydrodynamic fields which depend only on the radial coordinate. For the one-dimensional case we then provide the complete solution. 

Dimensional analysis alone gives the position of the shock wave $R=R(t)$ in terms of time $t$, the released energy $E$, and the background density $\rho_\infty$:
\begin{equation}
\label{R-d}
R(t)= \left(\frac{Et^2}{A \rho_\infty}\right)^\frac{1}{d+2}
\end{equation}
The determination of the amplitude $A$ requires an effort~\cite{Sedov2014,ZeldovichBook,LandauBook,BarenblattBook},  the dependence of the position of the shock $R(t)$ on the basic parameters automatically follows from dimensional considerations.

The density is uniform and it equals to $\rho_\infty$ everywhere in front of the shock wave, i.e., for $r>R(t)$. The velocity of the shock wave is
\begin{equation}
\label{U-d}
U = \frac{dR}{dt} = \delta\,\frac{R}{t}, \qquad \delta\equiv \frac{2}{d+2}
\end{equation}

Behind the shock wave, $0\leq r<R(t)$,  the radial velocity $v(r,t)$, density $\rho(r,t)$ and pressure $p(r,t)$ satisfy 
\begin{subequations}
\begin{align}
\label{cont-eq}
&\partial_t \rho + \partial_r (\rho v) +\frac{d-1}{r}\,\rho v  = 0\\
\label{E-eq}
&(\partial_t + v \partial_r) v +\frac{1}{\rho}\, \partial_r P = 0 \\
\label{entropy-eq}
&(\partial_t + v \partial_r)\ln\frac{P}{\rho^\gamma} =0
\end{align}
\end{subequations}
where $\gamma$ is the adiabatic index. These equations follow from the Euler equations for the conserved fields and use of the entropy form for the ideal gas [see App.~(\ref{app:idealgas})].
The Rankine-Hugoniot conditions  \cite{LandauBook} describing the jump between the states on both sides of the shock wave simplify, in the case of the infinitely strong blast, to 
\begin{equation}
\label{RH}
\frac{\rho(R)}{\rho_\infty}=  \frac{\gamma+1}{\gamma-1}\,, \quad 
\frac{v(R)}{U}=\frac{2}{\gamma+1}\,,\quad 
\frac{P(R)}{\rho_\infty U^2} = \frac{2}{\gamma+1}.
\end{equation}
The pressure $p_\infty$ in front of the shock wave can be neglected if $p(R)\gg p_\infty$. Using Eqs.~\eqref{R-d}--\eqref{U-d} and \eqref{RH} one finds that this is valid in the time range  
\begin{equation}
t\ll t_*, \quad  t_* = \left(\frac{E}{p_\infty}\right)^\frac{1}{d}\sqrt{\frac{\rho_\infty}{p_\infty}}
\end{equation}
In our case of a quiescent gas, $p_\infty=0$, so $t_* = \infty$ and the blast forever remains infinitely strong.

Instead of pressure, we can use the temperature field given by (for the ideal gas) $T= \mu P/\rho$. The dimensional analysis assures that the hydrodynamic variables acquire a self-similar form 
\begin{equation}
\label{scaling}
 \rho=\rho_\infty G(\xi),\quad v=\delta\,\frac{r}{t}\,V(\xi),  \quad T=\f{\mu \delta^2}{\gamma}\,\frac{r^2}{t^2}\,Z(\xi).
\end{equation}
This differs from Eqs.~(\ref{TvNS1}-\ref{TvNS3}) by  factors involving $\delta,\gamma$ which are inserted for convenience --- e.g from Eq.~\eqref{U-d} we see that the velocity of the shock wave is $\delta {R}/{t}$ and this suggests the use of the factor $\delta$ for $v$. The fields now depend on the single dimensionless  variable
\begin{equation}
\label{xi-def}
\xi = \frac{r}{R}.
\end{equation}
One seeks the behavior of $G(\xi), V(\xi)$ and $Z(\xi)$ behind the shock wave, $0\leq \xi\leq 1$. The Rankine-Hugoniot conditions Eq.~\eqref{RH} become
\begin{subequations}
\begin{align}
\label{RH-G}
& G(1)= \frac{\gamma+1}{\gamma-1},\\
\label{RH-V}
& V(1)=\frac{2}{\gamma+1},\\
\label{RH-Z}
& Z(1)=\frac{2\gamma(\gamma-1)}{(\gamma+1)^2}.
\end{align}
\end{subequations}
The conservation of energy allows to express $Z$ through the scaled velocity $V$:
\begin{equation}
\label{ZV}
Z = \frac{\gamma(\gamma-1)(1-V)V^2}{2(\gamma V -1)}.
\end{equation}
This integral of motion is usually established~\cite{LandauBook}  for $d=3$, but the same derivation works in arbitrary dimension and yields the universal result Eq.~\eqref{ZV}. 

Plugging the ansatz Eqs.~\eqref{scaling}--\eqref{xi-def} into Eq.~\eqref{cont-eq} we obtain
\begin{equation}
\label{VG-eq}
\frac{dV}{d\ell}+(V-1)\,\frac{d\ln G}{d\ell}    = - dV,  
\end{equation}
where $\ell=\ln\xi$. Similarly we transform Eq.~\eqref{entropy-eq} into
\begin{equation}
\label{ZG-eq}
\frac{d\ln Z}{d\ell} - (\gamma-1)\,\frac{d\ln G}{d\ell} = \frac{d+2-2V}{V-1}.
\end{equation}

Equations \eqref{VG-eq}--\eqref{ZG-eq} are solvable for arbitrary $d$ and $\gamma>1$. One must solve these equations even if one merely wants to determine the amplitude $A=A(d,\gamma)$ in Eq.~\eqref{R-d}. The energy conservation gives
\begin{eqnarray}
\label{energy}
E  &=& \int_0^R dr\,\,\Omega_d\, r^{d-1}\,\rho\left[\frac{v^2}{2}+\frac{c^2}{\gamma(\gamma-1)}\right] \nonumber \\
    &=& \rho_\infty\,\Omega_d\, \delta^2\,\frac{R^{d+2}}{t^2} \int_0^1 d\xi\,\xi^{d+1}\,\frac{(\gamma-1)V^3}{2(\gamma V -1)}\,G,
\end{eqnarray}
where $\Omega_d$ is the surface area of the unit sphere and we have used Eq.~\eqref{ZV}. Combining Eq.~\eqref{energy} with Eq.~\eqref{R-d} we obtain
\begin{equation}
\label{A-int}
A =\Omega_d\, \delta^2\,\frac{\gamma-1}{2} \int_0^1 d\xi\,\xi^{d+1}\,\frac{V^3}{\gamma V -1}\,G.
\end{equation}

In the classical literature, the blast problem is studied in three dimensions; the two-dimensional solutions are mentioned in \cite{Sedov2014}. Hence we now present the derivation of the solution in $d=1$ which we need to compare with results of direct microscopic simulations. The problem is solvable for arbitrary adiabatic index $\gamma>1$, but since we consider the monoatomic gas, we use $\gamma=3$ following from the general prediction of kinetic theory, $\gamma=1+2/d$ for monoatomic gases. The Rankine-Hugoniot conditions Eqs.~\eqref{RH-G}--\eqref{RH-Z} become
\begin{equation}
\label{BC}
G(1) = 2, \quad V(1) = \tfrac{1}{2}, \quad  Z(1) = \tfrac{3}{4}.
\end{equation}
We insert Eq.~\eqref{ZV} into  Eq.~\eqref{ZG-eq} and find
\begin{eqnarray}
\label{ZG-eq-d}
&&\left[\frac{1}{V}+\frac{1}{V-1}-\frac{3}{3V -1}\right]\frac{dV}{d\ell} \nonumber \\
&& ={2}\,\frac{d\ln G}{d\ell} + \frac{3-2V}{V-1}.
\end{eqnarray}
Using Eqs.~\eqref{VG-eq} and \eqref{ZG-eq-d}, we express the derivatives of $V$ and $\ln G$ through $V$:
\begin{subequations}
\begin{align}
\label{V-diff}
&  2\,\frac{dV}{d\ell}  = - V\,\frac{3-13 V + 12 V^2}{1-4V+6 V^2},       \\
\label{G-diff}
& 2\,\frac{d\ln G}{d\ell} = V\,\frac{5V - 1}{1-5V+10 V^2 - 6 V^3}.
\end{align}
\end{subequations}
Dividing Eq.~\eqref{G-diff} by Eq.~\eqref{V-diff} yields
\begin{equation}
\frac{d\ln G}{d V} = \frac{5V-1}{(3V-1)(3-4V)(1-V)}
\end{equation}
which is integrated to give
\begin{subequations} 
\begin{equation}
\label{GV:1}
G = 2^\frac{16}{5}\, (1-V)^2\, (3V-1)^\frac{1}{5}\, (3-4V)^{-\frac{11}{5}}.
\end{equation}
The amplitude is fixed by the boundary conditions Eq.~\eqref{BC}. Similarly integrating Eq.~\eqref{V-diff} one gets
\begin{equation}
\label{xi-V:1}
\xi^5 = 2^{-\frac{4}{3}}\,(3V-1)^2\, V^{-\frac{10}{3}}\,(3-4V)^{-\frac{11}{3}}
\end{equation}
that implicitly determines $V=V(\xi)$. Finally $Z$ is given by Eq.~\eqref{ZV} which when $\gamma=3$ becomes 
\begin{equation}
\label{ZV:1}
Z = \frac{3(1-V)V^2}{3V - 1}.
\end{equation}
\end{subequations}
Equations \eqref{GV:1}--\eqref{ZV:1} constitute the exact solution. 
\begin{figure*}
	\begin{center}
		\leavevmode
		\includegraphics[width=5.5cm,angle=0]{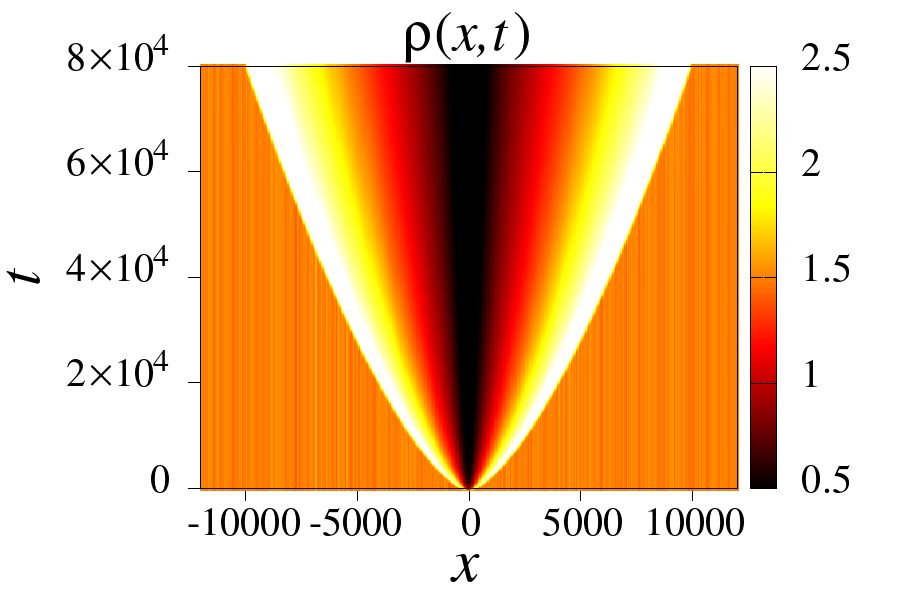}
		\includegraphics[width=5.5cm,angle=0]{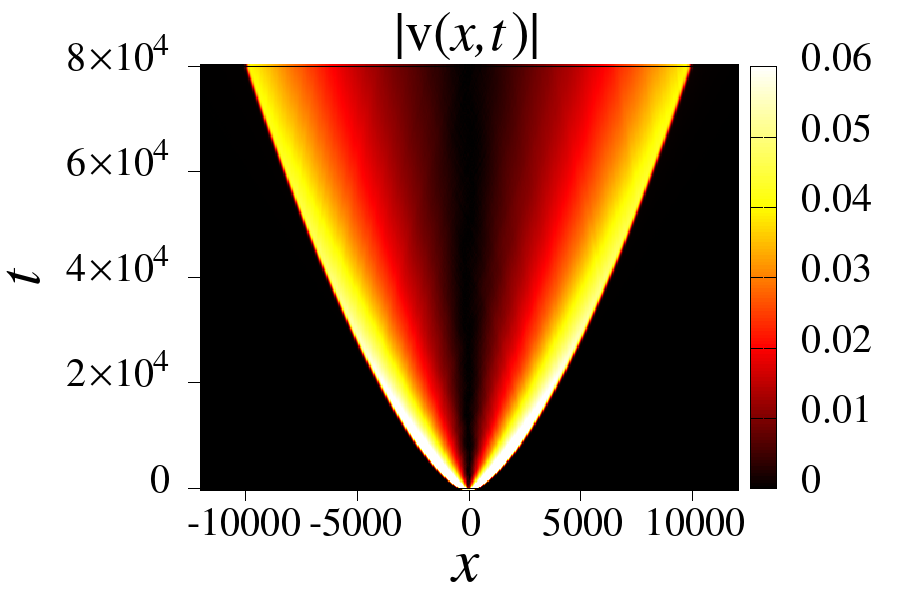}
		\includegraphics[width=5.5cm,angle=0]{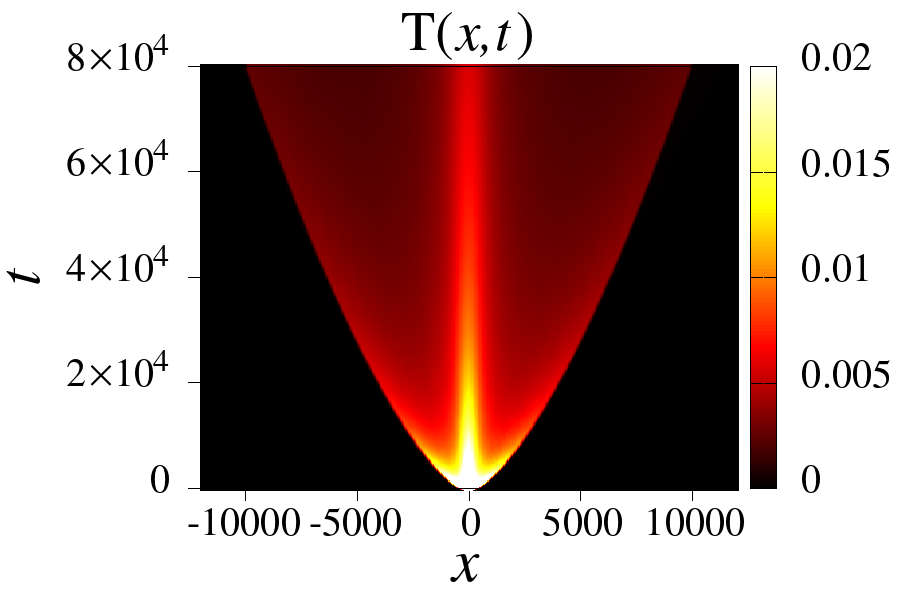}
		\caption{{\it Heat maps showing the spatio-temporal evolution of the  density, velocity and temperature fields, starting from initial conditions corresponding to a Gaussian initial temperature profile and $\rho(x,0)=\rho_\infty=1.5, v(x,0)=0$. The simulation parameters were $N=L=24000$, $E =32, \mu=1.5$ and an ensemble average over $10^4$ initial conditions were performed.}}
		\label{heatmap}
	\end{center}
\end{figure*}
When $d=1$, we have $\Omega_1=2, ~\delta=2/3$ and $\gamma=3$. Therefore Eq.~\eqref{A-int} reduces to 
\begin{equation}
\label{A:int-1}
A = \left(\frac{2}{3}\right)^2\int_0^1 d\xi\,\xi^2\,\frac{2V^3}{3V-1}\,G
\end{equation}
Using Eqs.~\eqref{GV:1}--\eqref{xi-V:1} one can reduce the integral in Eq.~\eqref{A:int-1} to a rather complicated integral over $V$ which is computed to give
\begin{equation}
\label{A-1}
A=\frac{152}{1071}
\end{equation}
 Our basic analytical prediction about the position of the shock wave becomes 
\begin{equation}
\label{R-1}
R(t)= \left(\frac{1071}{152}\, \frac{Et^2}{\rho_\infty}\right)^\frac{1}{3}.
\end{equation}
This, along with Eqs.~(\ref{xi-def}, \ref{GV:1}, \ref{xi-V:1}, \ref{ZV:1}) provides the complete and explicit TvNS solution. In the next section we present a comparison of the TvNS solution with results from direct simulations of the microscopic model.

\section{Results of microscopic simulations and comparison with the TvNS solution}
\label{sec:micro}

We now present the results of the microscopic simulations. We will mainly discuss  the results of the Gaussian initial temperature profile and with the averaging over an initial ensemble of microstates where the total energy and total momentum are fixed exactly [ensemble (B) in previous section].  We also show a comparison with the empirical  profiles (obtained from spatial coarse-graining of single realization). We will then briefly mention results obtained for the box profile and for the ensemble choice where energy and momentum are fixed on average.

In Fig.~\eqref{heatmap} we show the spatio-temporal evolution of the  three fields $\rho(x,t)$, $v(x,t)$,  and $T(x,t)$ starting from a blast wave initial condition with a flat mass density, zero velocity field and a Gaussian temperature profile. We see a clear sub-ballistic evolution of the fields.  The mass density and flow velocity of the gas are peaked around the blast front while the temperature has an additional peak at the centre.

\begin{figure*}
	\begin{center}
		\leavevmode
		\includegraphics[width=5.5cm,angle=0]{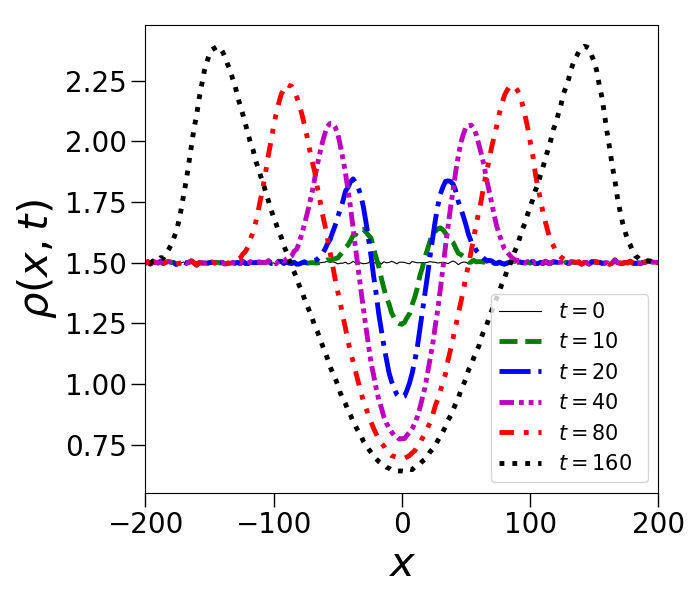}
		\includegraphics[width=5.5cm,angle=0]{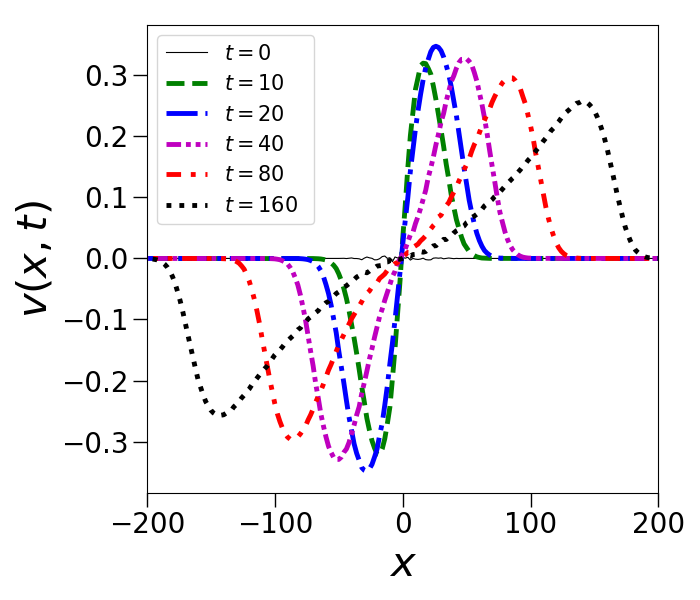}
		\includegraphics[width=5.5cm,angle=0]{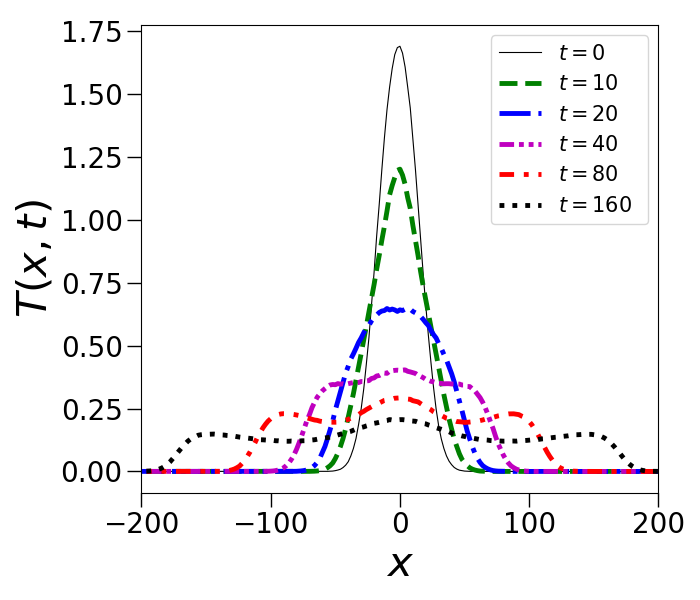}
		\caption{{\it Early time evolution of density, velocity and temperature fields, obtained by ensemble averaging, starting from initial conditions corresponding to a Gaussian initial temperature profile and $\rho(x,0)=\rho_\infty=1.5, v(x,0)=0$. The simulation parameters were $N=L=512$, $E =32, \mu=1.5$ and an average over $10^5$ initial conditions were performed.}}
		\label{earlytime}
	\end{center}
\end{figure*}

\begin{figure*}
	\begin{center}
		\leavevmode
		\includegraphics[width=5.5cm,angle=0]{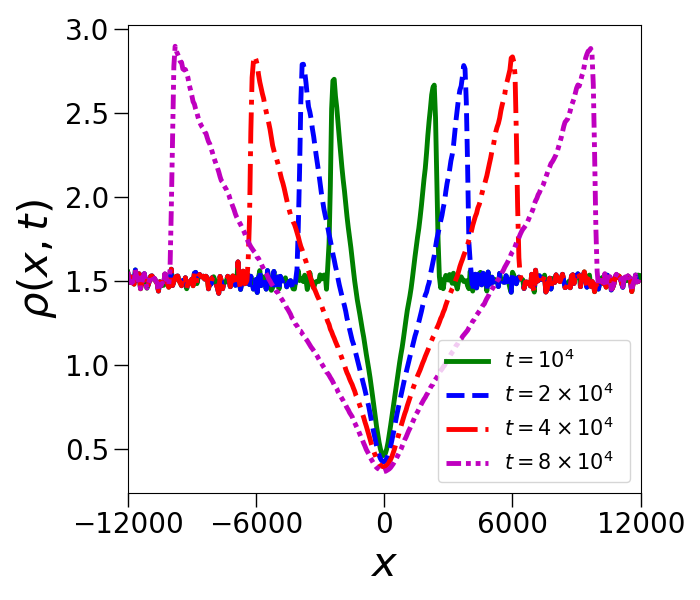}
		\includegraphics[width=5.5cm,angle=0]{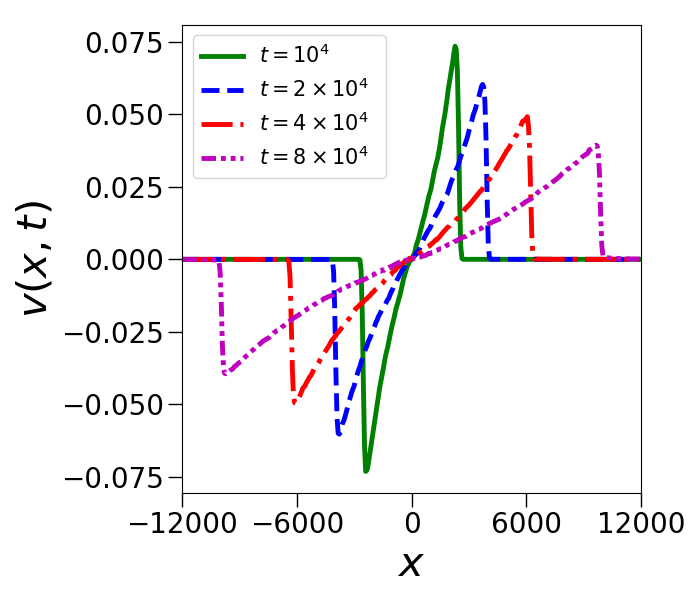}
		\includegraphics[width=5.5cm,angle=0]{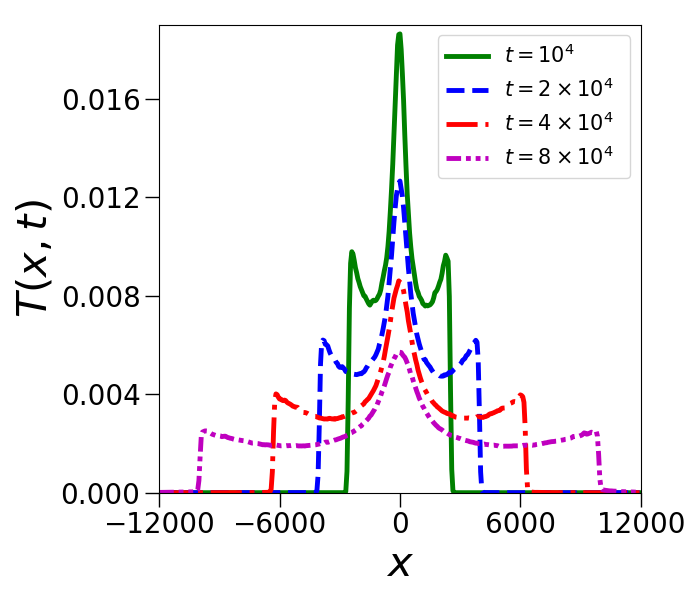}
		\includegraphics[width=5.5cm,angle=0]{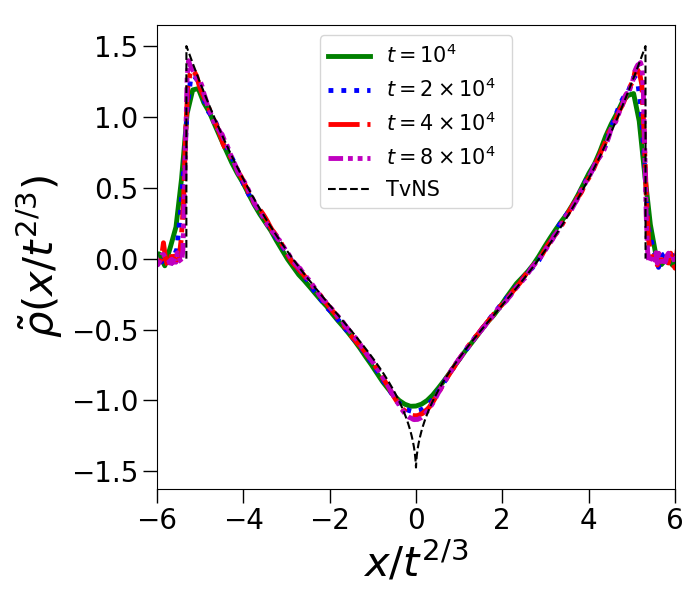}
		\includegraphics[width=5.5cm,angle=0]{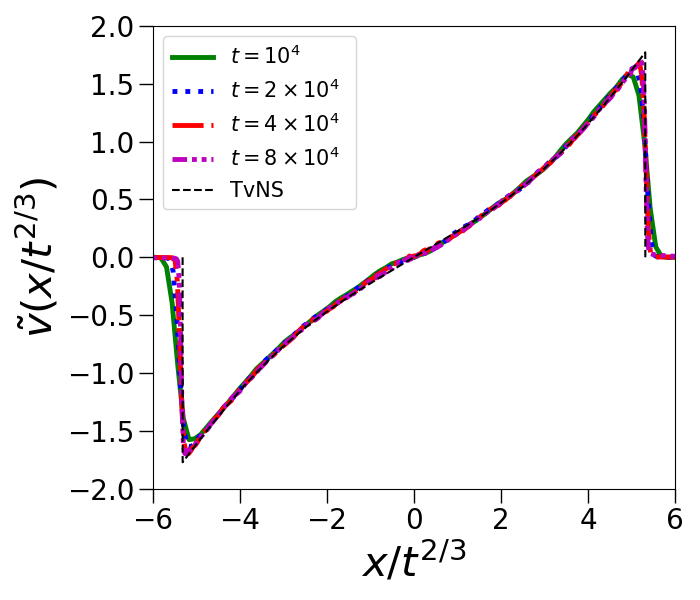}
		\includegraphics[width=5.5cm,angle=0]{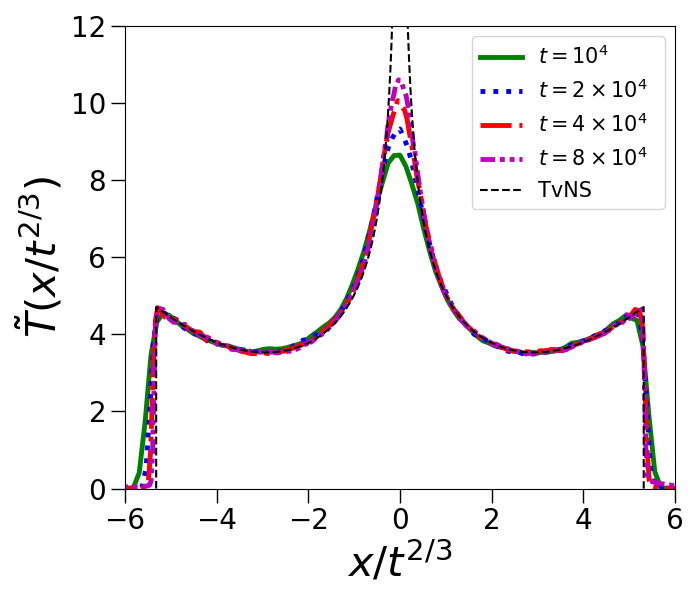}
		\caption{{\it (Top panel) Late time evolution of density, velocity and temperature fields, starting from the initial conditions corresponding to a Gaussian initial temperature profile and $\rho(x,0)=\rho_\infty=1.5, v(x,0)=0$. The simulation parameters were $N=L=24000$, $E =32, \mu=1.5$ and an average over $10^4$ initial conditions were performed. (Lower panel) This shows the $x/t^{2/3}$ scaling of the data. We observe an excellent collapse of the data at the longest times everywhere except near the origin.  The exact scaling solution of the hydrodynamic Euler equations is shown by the black dashed line.}}
		\label{longtime}
	\end{center}
\end{figure*}

In Fig.~(\ref{earlytime}) we plot the early time evolution of the three fields $\rho(x,t)$, $v(x,t)$,  and $T(x,t)$ as  functions of position $x$ at different times $t$. In Fig.~(\ref{longtime}) we plot the  evolution of the blast wave at long times. The plots in the lower panel in  Fig.~(\ref{longtime}) show the scaled fields $\widetilde{\rho}=\rho_\infty G-\rho_\infty,~\widetilde{v}=t^{1/3} v(x,t) =(2 \alpha/3) \xi V$ and $\widetilde{T}=t^{2/3} T(x,t)=(4 \mu \alpha^2/27) \xi^2 Z$, where  $\alpha = [E/(A \rho_\infty)]^{1/3}$, as  functions of the scaling variable $x/t^{2/3}$. 
We find a very good scaling collapse of the data confirming the expected TvNS scaling. For comparison, we plot the exact predictions of  TvNS solution given by Eqs.~(\ref{GV:1}, \ref{xi-V:1}, \ref{ZV:1}) (black dashed lines in the scaled plots). We find a near perfect agreement between the simulation results and the TvNS solution except for some deviations at the centre of the blast. As we discuss in the next section, the TvNS solution in fact predicts a divergence of the temperature field with $\widetilde{T}(\xi) \sim \xi^{-1/2}$ and this implies that the effect of the heat conduction term in the hydrodynamic equation for energy conservation cannot be neglected. Thus one has to deal with the full Navier-Stokes-Fourier equations instead of the Euler equations.  
In the next section we discuss in  detail the behaviour of the fields in the central region and compare the simulation results with the solution of the full hydrodynamic equations.

\begin{figure*}
%	\begin{center}
%		\leavevmode
\centering
		\includegraphics[width=5.5cm,angle=0]{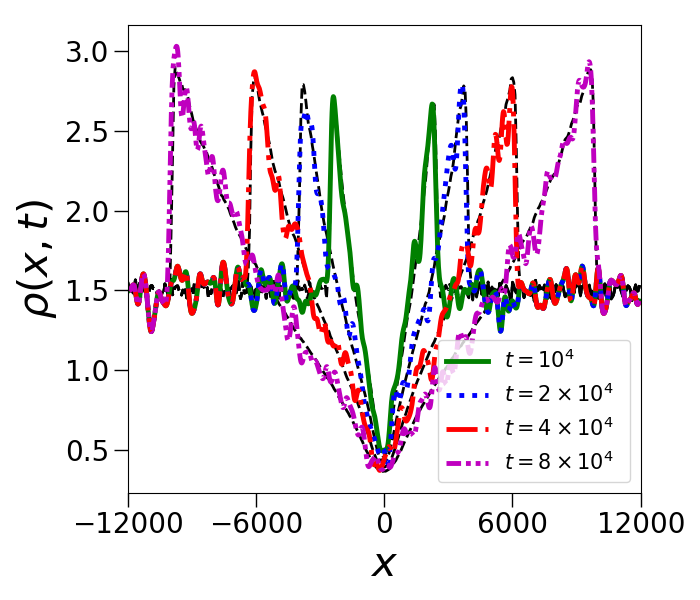}
		\includegraphics[width=5.5cm,angle=0]{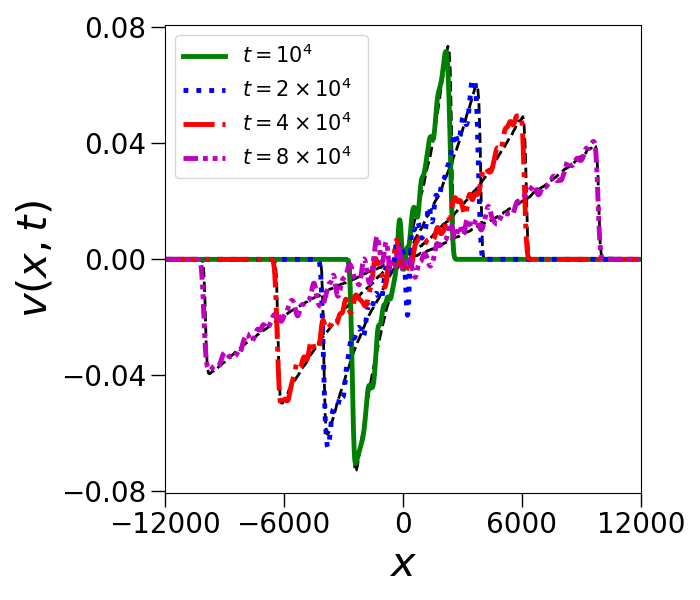}
		\includegraphics[width=5.5cm,angle=0]{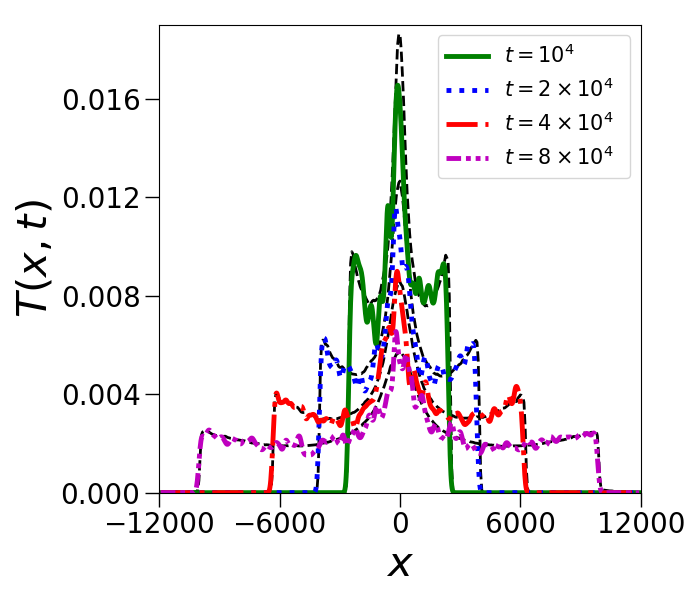}
		\caption{{\it Evolution of the empirical density, velocity and temperature fields, as obtained from  a single realization, and after spatial coarse graining according to Eqs.~(\ref{br}-\ref{bE}).  The black dashed lines indicate the ensemble averaged fields and we see that apart from the flutuations seen in the empirical fields, there is very good agreement between the two. The parameter values for this simulation were $N=L=24000$ and a coarse-graining length scale,  $\ell=50$, and same initial conditions as in Fig.~\eqref{longtime}. %\textcolor{blue}{(Lower panel) $x/t^{2/3}$ scaling of the empirical fields. The exact scaling solution of the hydrodynamic Euler equations is shown by the black dashed line.}
}}
		\label{noscalespace}
%	\end{center}
\end{figure*}

{\bf Comparison with the empirical fields}:
Ideally we would like to see a comparison of the hydrodynamic theory with simulations for the empirical fields which corresponds to the physical situation where we observe a {\it{single}} realization and there is only a spatial averaging of the microscopic degrees. The empirical  fields are defined in Eqs.~(\ref{br},\ref{bp},\ref{bE}) and we now present the results of an average over a spatial length scale of $\ell=50$. The system is initially excited in the same manner as the previous subsection with the same initial energy $E $ and zero momentum, but we now take only one realization of the microscopic process.  The comparison  of the empirical density profiles with the ensemble averaged profiles is shown in Fig.~(\ref{noscalespace}) and we see close agreeement between the two. It is expected that the fluctuations seen in the empirical profile should vanish in the thermodynamic limit. 

\begin{figure*}
	\begin{center}
		\leavevmode
	        \includegraphics[width=5.5cm,angle=0]{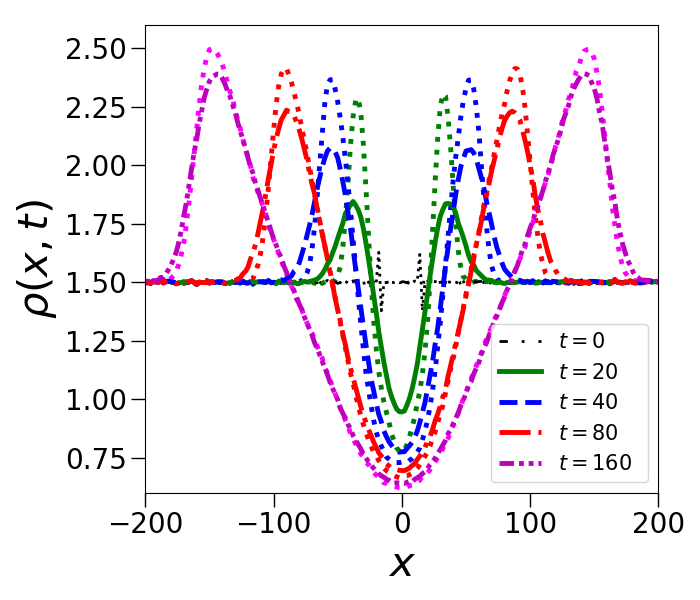}
		\includegraphics[width=5.5cm,angle=0]{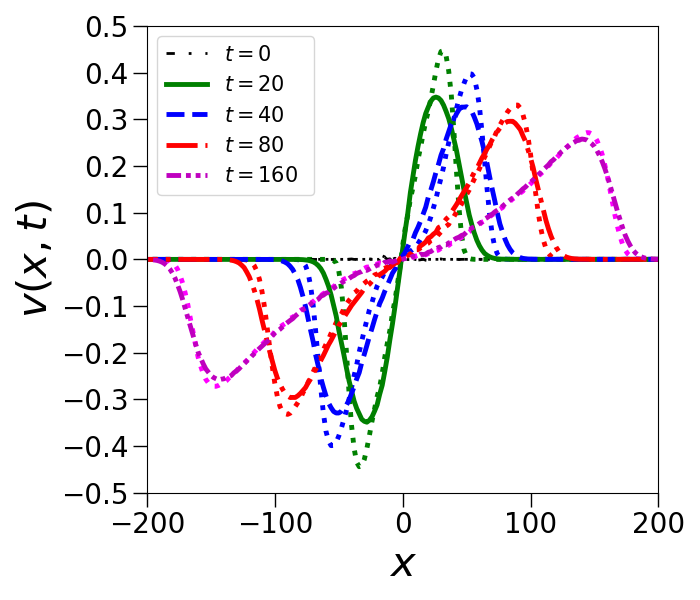}
		\includegraphics[width=5.5cm,angle=0]{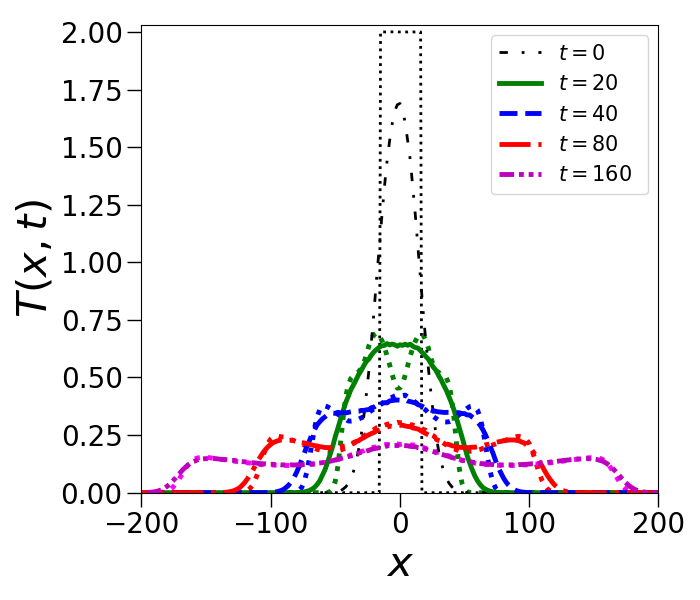}
		\caption{{\it This plot shows a comparison of the  evolution of three field profiles starting from two different initial conditions corresponding to either a Gaussian temperature profile (solid lines) or a box-profile (dotted lines). The total energy was fixed at $E =32$,  the background density was taken as $\rho_\infty=1.5$ and the average was taken over $10^5$ realizations.}}
		\label{ICcomp}
	\end{center}
\end{figure*}

{\bf Dependence on initial conditions}: It is expected that the long-time scaling solution should be independent of the details of the initial conditions and depends on just the two parameters, namely the total blast energy, $E $, and the background density $\rho_\infty$. We verify this in our simulations by starting with the box initial condition discussed in Eq.~\eqref{eq:ezero}, with the same values of $E =32$ and $\rho_\infty=1.5$. From the results plotted in Fig.~\eqref{ICcomp}, it is clear that the profiles of all the three fields from the two different initial conditions quickly converge and become almost indistinguishable. 

\begin{figure*}
	\begin{center}
		\leavevmode
		\includegraphics[width=5.5cm,angle=0]{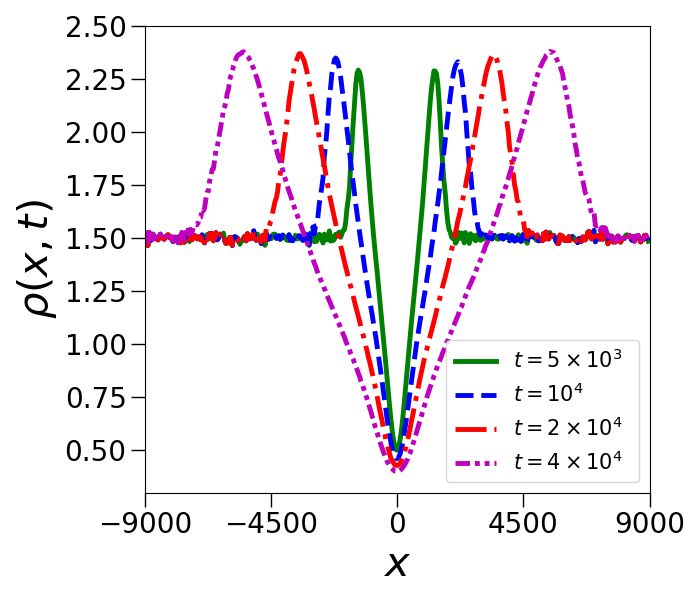}
		\includegraphics[width=5.5cm,angle=0]{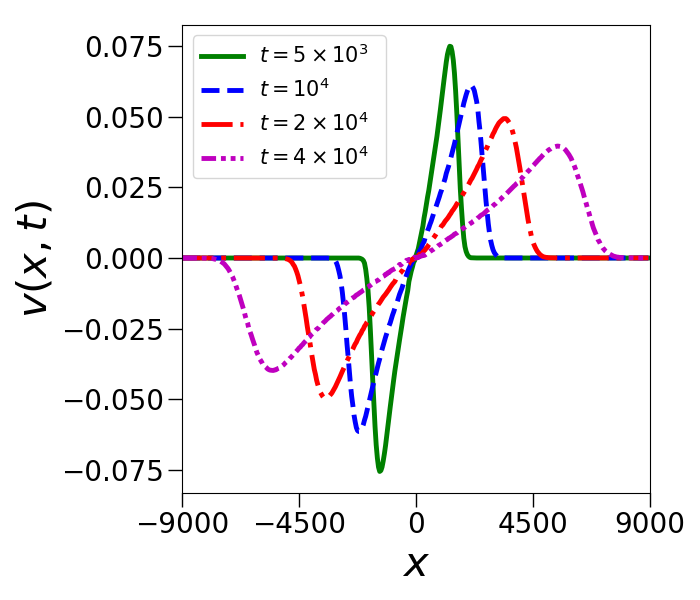}
		\includegraphics[width=5.5cm,angle=0]{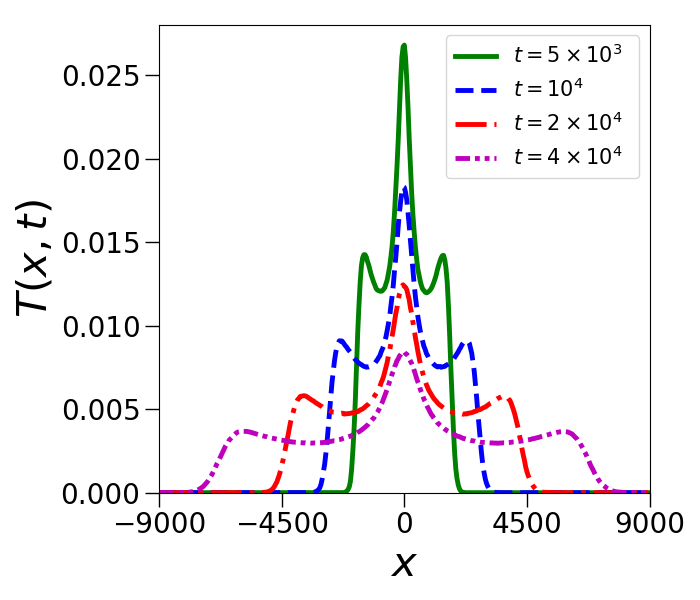}
		\includegraphics[width=5.5cm,angle=0]{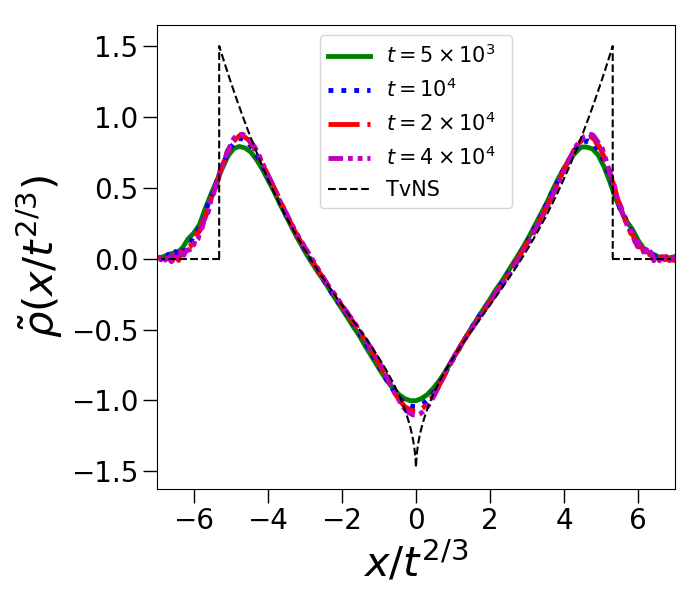}
		\includegraphics[width=5.5cm,angle=0]{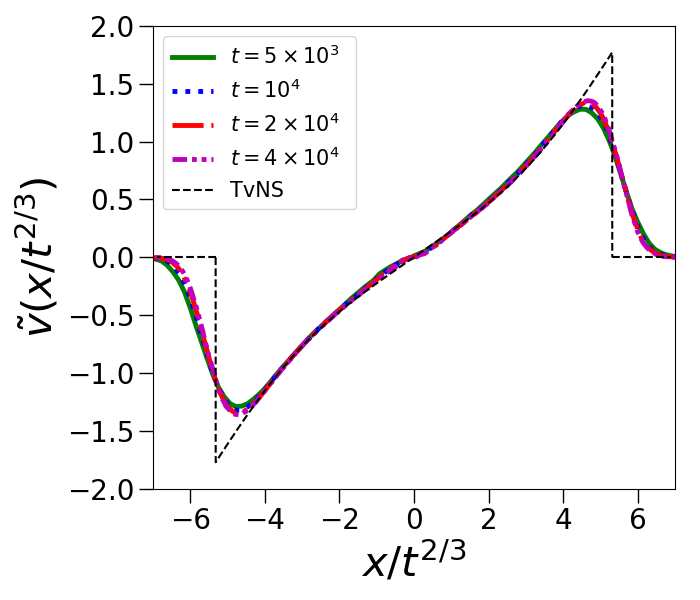}
		\includegraphics[width=5.5cm,angle=0]{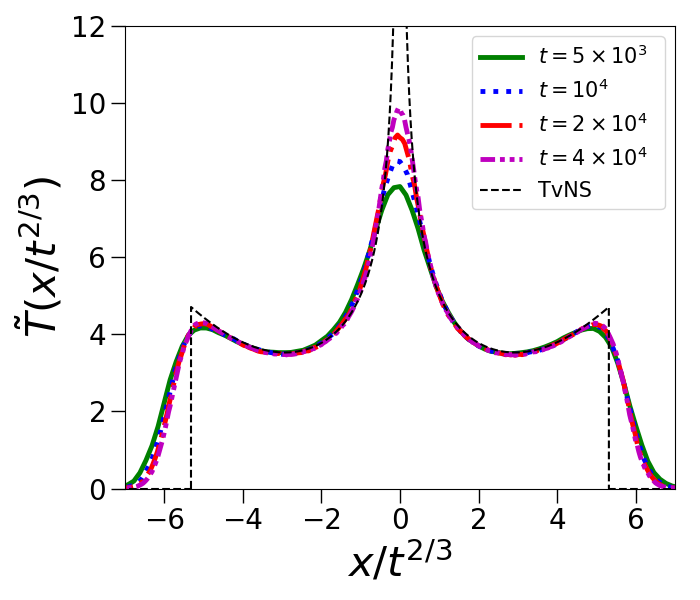}
		\caption{{\it {\bf Initial ensemble with fluctuations in initial total energy and total momentum, with mean energy fixed at $E $ and mean momentum at $0$} (Top panel) Evolution of density, velocity and temperature fields, starting from the initial conditions corresponding to a Gaussian initial temperature profile and $\rho(x,0)=\rho_\infty=1.5, v(x,0)=0$. The simulation parameters were $N=L=18000$, $E =32, \mu=1.5$ and an average over $10^4$ initial conditions were performed. (Lower panel) This shows the $x/t^{2/3}$ scaling of the data. We again observe a collapse of the data at the longest times, everywhere except at the centre. However, comparing with Fig.~\eqref{longtime} we see that the agreement with the exact TvNS scaling solution, shown by dashed lines, is now much poorer with significant deviations both at the blast centre as well as near the shock front. }}
		\label{longtimecan}
	\end{center}
\end{figure*}

{\bf Dependence on initial ensemble}: In  Fig.~\eqref{longtimecan} we repeat the simulations with initial conditions chosen from the ensemble~(A) discussed in Sec.~\eqref{sec:model} where the energy $E $ and total zero initial momentum value are fixed only on average. This means that there are realizations where the energy can be larger or smaller than $E $ while the momentum can have a non-zero value. As a result, this leads to tails developing in the shock front and we lose the agreement with the hydrodynamic theory. The correct physically relevant initial condition is the fixed energy, fixed momentum ensemble and so we will not discuss further the case of ensemble~(A). 

\section{The core region: results from the Navier-Stokes-Fourier equations}
\label{sec:hydro}
We  now  compare our results for the evolution of the conserved fields  with those from solving the full hydrodynamic equations for the one-dimensional fluid. These are the one-dimensional Navier-Stokes-Fourier equations for the fields $\rho,v$ and $e$:
\begin{subequations}
\begin{align}
&\partial_t \rho + \partial_x (\rho v) = 0\label{tNS1} \\
&\partial_t (\rho v) + \partial_x (\rho v^2 +P  ) = \p_x (\zeta\partial_x v) \label{tNS2} \\
&\partial_t (\rho e) + \partial_x (\rho ev   + P v)= \p_x (\zeta v\partial_x v + \kappa\partial_x T ),
\label{tNS3}
\end{align}
\end{subequations}
where $\zeta$ denotes the bulk viscosity and $\kappa$ is the thermal conductivity
of the system. These transport coefficients can depend on the fields and, based on  Green-Kubo relations as well as kinetic theory arguments~\cite{krapivskybook}, we expect their temperature dependence to be $\zeta \sim T^{1/2}$ and $\kappa \sim T^{1/2}$. A recent numerical study \cite{Hurtado2016} suggests the density dependence $\kappa \sim \rho^{1/3}$. In our numerical study we have thus used the forms $\zeta =D_1T^{1/2}$ and $\kappa=D_2\rho^{1/3}T^{1/2}$, where $D_1$ and $D_2$ are constants. Note that the above equations need to be supplemented with knowledge of thermodynamics properties of the system as given in Eq.~\eqref{therm1} and Eq.~\eqref{therm2}.  We will first discuss results obtained from a direct numerical solution of the hydrodynamic equations. In the subsequent subsection we will present a more detailed discussion of the self-similar form and the related scaling functions.
\begin{figure*}
	\begin{center}
		\leavevmode
		\includegraphics[width=5.5cm,angle=0]{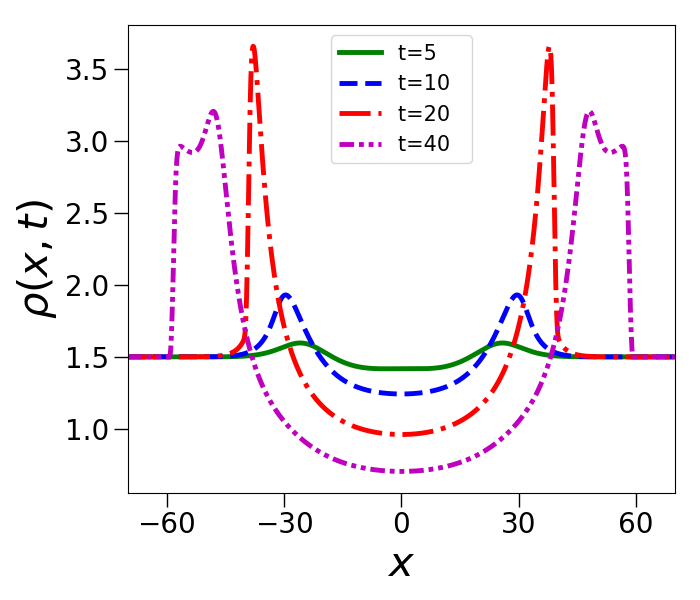}
		\includegraphics[width=5.5cm,angle=0]{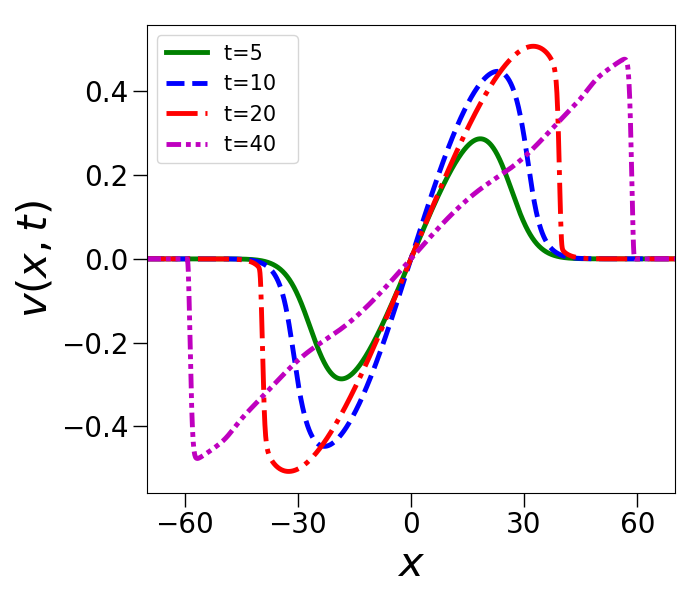}
		\includegraphics[width=5.5cm,angle=0]{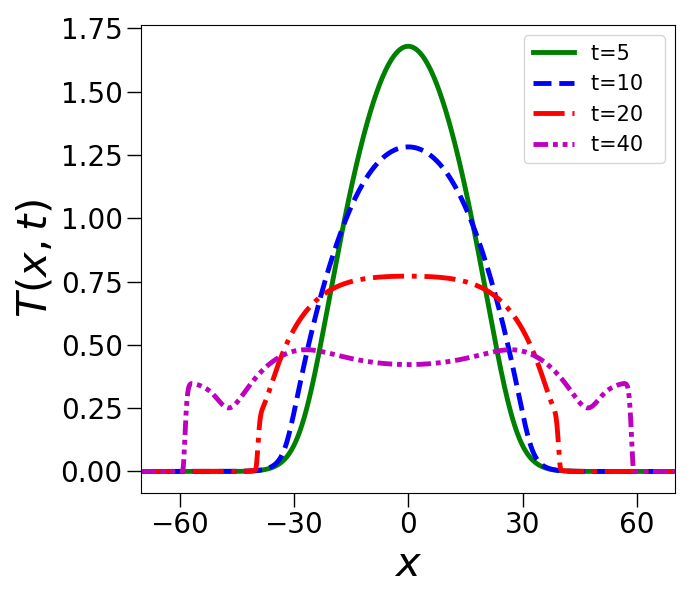}
		\caption{{\it  Early time evolution of the hydrodynamic fields $\rho(x,t), v(x,t), T(x,t)$ as obtained from a numerical solution of Eqs.~(\ref{tNS1}-\ref{tNS3}), starting from the  initial conditions with Gaussian temperature profile used  in the simulations for Fig.~\eqref{ICcomp}. We see the formation of a sharp front. The other parameters used in the numerics  are $D_1=1$, $D_2=1$ and $L=4000$.}}
		\label{hydroearly}
	\end{center}
\end{figure*}

\begin{figure*}
	\begin{center}
		\leavevmode
		\includegraphics[width=5.5cm,angle=0]{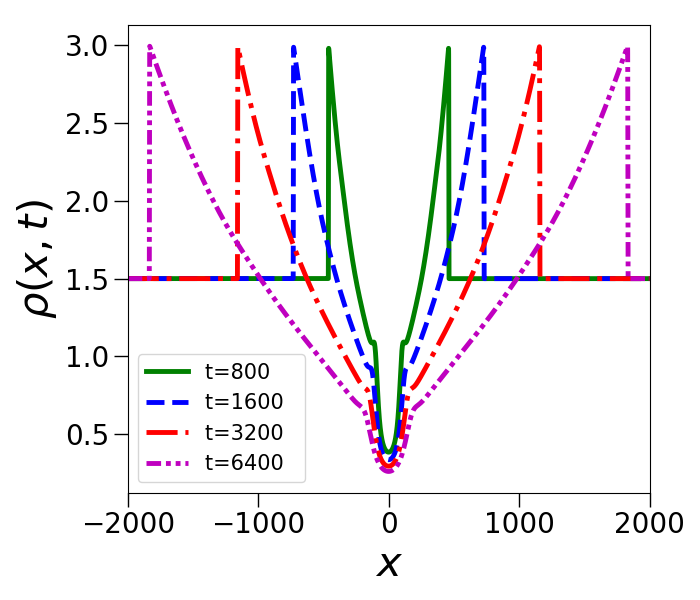}
		\includegraphics[width=5.5cm,angle=0]{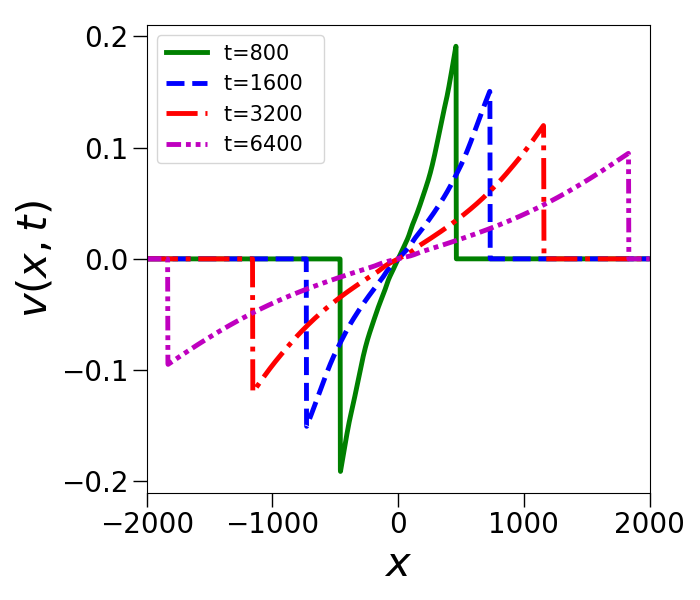}
		\includegraphics[width=5.5cm,angle=0]{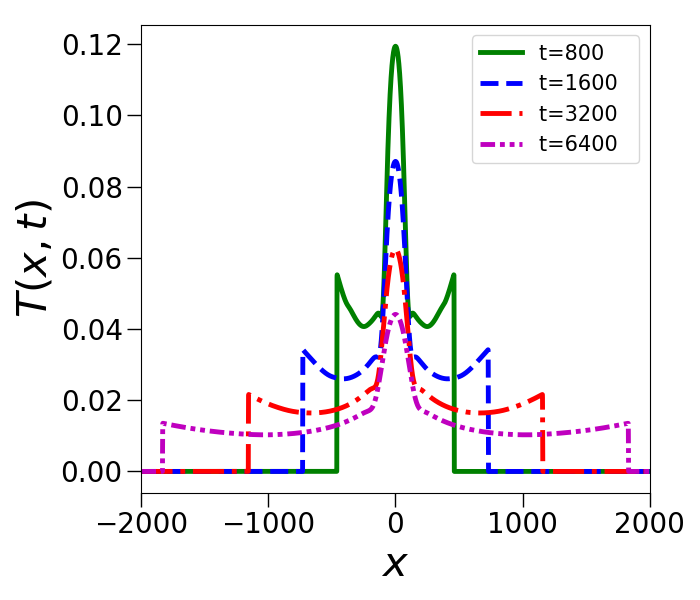}
		\includegraphics[width=5.5cm,angle=0]{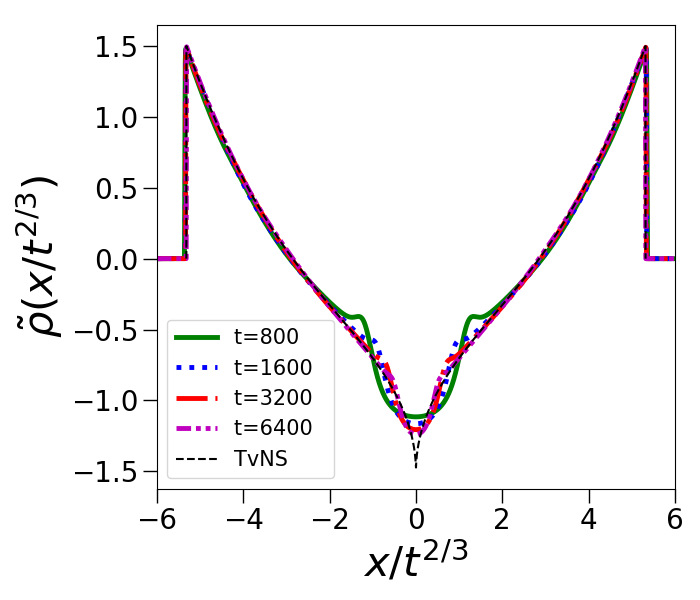}
		\includegraphics[width=5.5cm,angle=0]{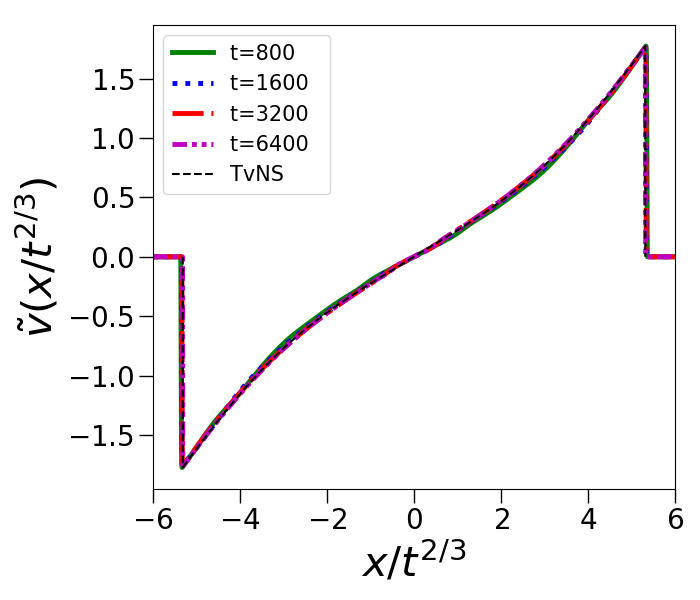}
		\includegraphics[width=5.5cm,angle=0]{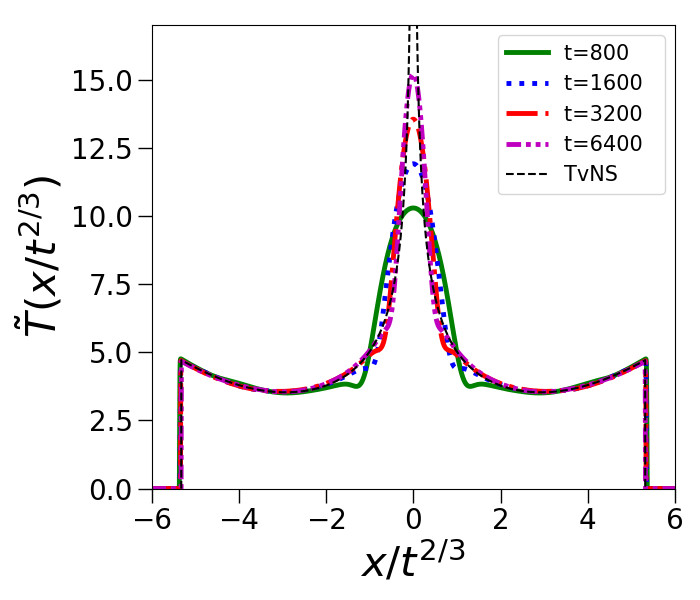}
		\caption{{\it (Top panel) Evolution of the hydrodynamic fields $\rho(x,t), v(x,t), T(x,t)$ as obtained from a numerical solution of Eqs.~(\ref{tNS1}-\ref{tNS3}), starting from the  initial conditions with Gaussian temperature profile used  in the simulations for Fig.~\eqref{ICcomp}.  (Lower panel) Plots of the scaling functions $\widetilde{\rho}, \widetilde{v}, \widetilde{T}$ and a comparison with the TvNS solution (black dashed lines). The other parameters used in the numerics  are $D_1=1$, $D_2=1$ and $L=4000$.}}
		\label{noscaleHydnew}
	\end{center}
\end{figure*}

%\begin{figure}
%	\begin{center}
%		\leavevmode
%		\includegraphics[width=6.5cm,angle=0]{mid_point_temperature.pdf}
%		\caption{{\it Plot of $T(0,t)$ for the same data as in Fig.~\eqref{longtime}. We see a logarithmic correction to the expected $t^{-2/3}$ behaviour.}}
%		\label{T0t}
%	\end{center}
%\end{figure}

\subsection{Numerical solution of hydrodynamic equations}
\label{sec:hydronumerics}

We now consider the evolution of the density fields as obtained from a numerical solution of the  hydrodynamic NSF equations Eqs.~(\ref{tNS1})-(\ref{tNS3}). We take the same initial conditions as considered in the microscopic simulations, namely $\rho(x,0)=\rho_\infty, v(x,0)=0$ and $E(x,0)$ given by the analytic form in Eq.~\eqref{Eerf}. The  numerical solution used the MacCormack method~\cite{maccormack1982}, which is second order in both space and time and used a discretization $dx=0.1$ and $dt=0.001$. The diffusion constants in the dissipative terms were set to the values $D_1=D_2=1$ and the system size was taken as $L=4000$.  We have  evolved up to time  $t=6400$ which is  before the energy reaches the boundary.

\begin{figure*}
	\begin{center}
		\leavevmode
		\includegraphics[width=5.5cm,angle=0]{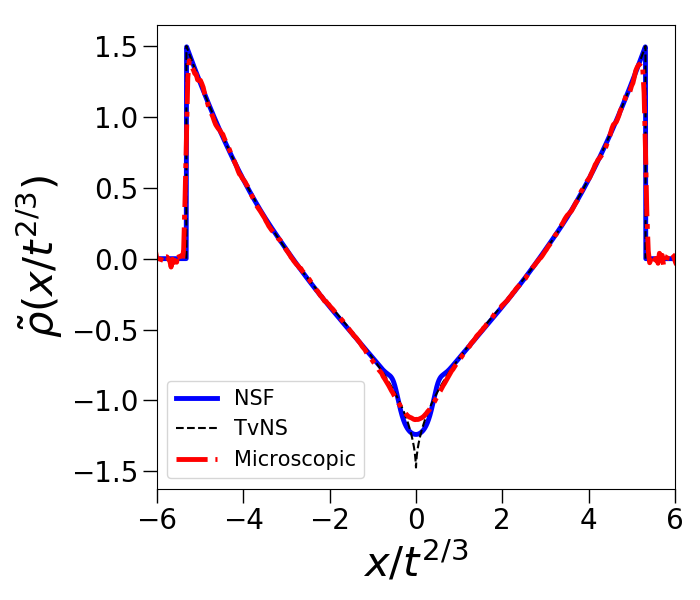}
		\includegraphics[width=5.5cm,angle=0]{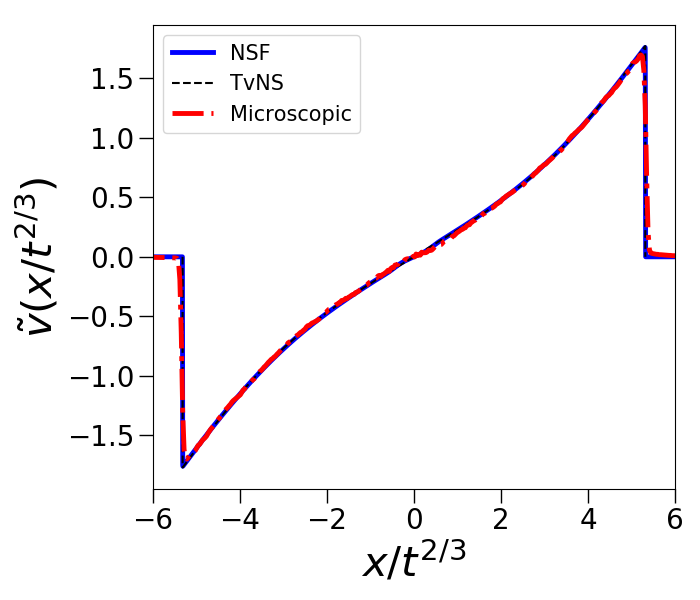}
		\includegraphics[width=5.5cm,angle=0]{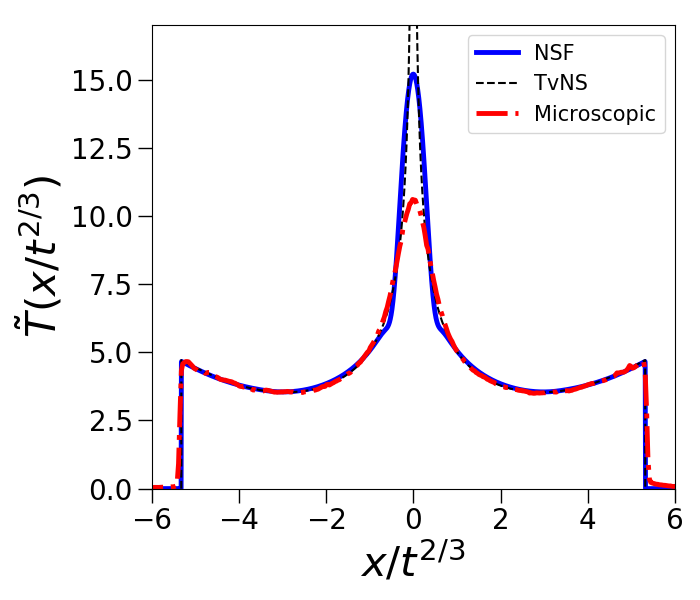}
		\caption{{\it Comparison of the scaled fields $\widetilde{\rho},\widetilde{v},\widetilde{T}$,  plotted as a function of the scaled variable $x/t^{2/3}$, from (a) the microscopic simulation (at time $t=80000$), (b) from dissipative Hydrodynamics (at $t=6400$) and (c) from  the exact TvNS solution of Euler equation.}}
		\label{comparisonlong}
	\end{center}
\end{figure*}

The numerical solution at early times is plotted in Fig.~\eqref{hydroearly} and we see that sharp fronts develop quickly for all the three fields. 
The late time numerical results are plotted in the top panel of Fig.~(\ref{noscaleHydnew}) where we show the profiles of the fields, $\rho(x,t)$, $v(x,t)$ and $T(x,t)$ at different times.  In  the lower panel of Fig.~(\ref{noscaleHydnew}) we verify the expected scaling forms  $\rho(x,t)= \rho_\infty+ \widetilde \rho(x/t^{2/3})$, $  v(x,t)= t^{-1/3} \widetilde v(x/t^{2/3})$, and $T(x,t)=t^{-2/3}\widetilde T(x/t^{2/3})$ but observe that there is again a core region where we do not get a good data collapse for the density and temperature fields.  For comparison, the exact TvNS scaling solution is  also plotted (dark dashed lines) and we see now that the singularities at the origin have now disappeared and the profiles are closer to those observed in simulations confirming thus the important role of dissipation.  
In Fig.~\eqref{comparisonlong} we show a comparison of the dissipative hydrodynamics numerical results (at the longest available time) with those from the MD simulations and from the exact TvNS solution of Euler equations.  It is clear from the results of simulations and the numerical solution of the NSF equations that the TvNS scaling breaks down in the core region of the blast and we should therefore look for a different scaling solution in this region. In the next section we analyze the NSF equations to answer this question.

%Finally we note that the hydrodynamic description assumes local thermodynamic equilibrium, and it fails inside the shock front where the hydrodynamic fields undergo large changes over very small length scale (comparable to mean free path generally, and to gaps between adjacent particles in one dimension). The Euler description where shock fronts are singular surfaces totally ignores the shock wave structure, while the NSF description is just an uncontrolled approximation. The description based on kinetic theory (the Boltzmann equation approach) is more adequate, but notoriously difficult for theoretical analysis even for infinitely strong shock waves~\cite{Cercignani1999,Takata2000}. For recent general discussion, see \cite{Colangeli2013}.

%% \begin{figure}
%% 	\begin{center}
%% 		\leavevmode
%% 		\includegraphics[width=4.cm,angle=0]{rhocomparisonshort.png}
%% 		\includegraphics[width=4.cm,angle=0]{velcomparisonshort.png}
%% 		\includegraphics[width=4.cm,angle=0]{enecomparisonshort.png}
%% 		\caption{{\it Comparison of the conserved quantities of the microscopic simulation and Hydrodynamics at $t=10$. We see a good agreement here.}}
%% 		\label{comparisonshort}
%% 	\end{center}
%% \end{figure}

\subsection{Analysis of the NSF equations and discussion of scaling in the blast core}
\label{sub-sec:origin}

Near the center of the explosion, the scaled hydrodynamic quantities exhibit the following behaviors 
\begin{subequations}
\begin{align}
\label{G-0}
 G & \simeq B\, |\xi|^\frac{1}{2}\,, ~\quad B =  2^\frac{16}{3}\,3^{-\frac{1}{2}}\,5^{-\frac{11}{6}} \\
\label{V-0}
 3V-1&\simeq \beta\, |\xi|^\frac{5}{2}\,, \quad ~~\beta = 2^\frac{2}{3}\,3^{-\frac{7}{2}}\,5^\frac{11}{6} \\
\label{Z-0}
 Z & \simeq C\, |\xi|^{-\frac{5}{2}}\,, \quad C =  2^\frac{1}{3}\,3^\frac{3}{2}\,5^{-\frac{11}{6}},
\end{align}
\end{subequations}
where $\xi=x/R(t)$. 
Indeed, Eq.~\eqref{V-0} follows from Eq.~\eqref{xi-V:1}. Substituting Eq.~\eqref{V-0} into Eq.~\eqref{GV:1} [resp. Eq.~\eqref{ZV:1}] one derives 
Eq.~\eqref{G-0} [resp. Eq.~\eqref{Z-0}].  

In one dimension, the scaling form in Eq.~\eqref{scaling} gives
\begin{equation}
\label{vrc}
\rho= \rho_\infty\, G(\xi),~~ v=\frac{2x}{3t}\, V(\xi), ~~T = \frac{4 \mu x^2}{27 t^2}\,  Z(\xi).
\end{equation}
These equations  are valid for all $x\in [-R(t), R(t)]$. The scaled variable is still given by $\xi=|x|/R$ and varies in the $0\leq \xi\leq 1$ range. Equations \eqref{G-0}--\eqref{Z-0} and \eqref{vrc} determine the behavior of the original hydrodynamic quantities near the center of explosion. Near the center of explosion, the gas moves with velocity varying linearly with distance from the center of explosion
\begin{equation}
\label{v-core}
v \simeq  \frac{2x}{9t}.
\end{equation}
Thus, the velocity vanishes at the origin. In contrast, the temperature diverges near the center of explosion. Using Eqs.~\eqref{Z-0} and \eqref{vrc} together with $T=\mu c^2/\gamma=\mu c^2/3$ one gets
\begin{equation}
T \simeq  \frac{4 \mu C}{27}\,\frac{x^2}{t^2}\,\xi^{-\frac{5}{2}}.
\end{equation}
Since $\xi\sim |x|/t^{2/3}$, the temperature diverges as
\begin{equation}
\label{outside}
T \sim |x|^{-\frac{1}{2}}\,t^{-\frac{1}{3}}.
\end{equation}
The divergence of the temperature is unphysical. From the simulation results of Sec.~\eqref{sec:micro}  we see  that the temperature at the center of the explosion remains finite  and actually decreases with time. The divergence is the consequence of the idealized modeling where we ignored the dissipative terms --- the divergence in fact indicates that the heat conduction term becomes non-negligible near the origin.  To rectify the analysis we now take into account the dissipative processes and consider an analysis of the full hydrodynamic equations given in Eq.~(\ref{tNS1}, \ref{tNS2}, \ref{tNS3}). As shown in App.~(\ref{app:core}) the terms corresponding to viscous dissipation drop off in the scaling limit being considered and we effectively get the following equations for the fields $\rho,v,T$:
\begin{subequations}
\begin{align} 
\label{cont-eq-1}
&\partial_t \rho + \partial_x (\rho v) = 0, \\
\label{NS-zeta}
&\rho (\partial_t + v \partial_x) v + \partial_x \f{\rho T}{\mu} = 0, \\
\label{heat}
&\frac{1}{2 \mu} \rho^3(\partial_t + v \partial_x)\left[\f{T}{\rho^2}\right] =  \partial_x \big(D_2T^{1/2}\,\rho^{1/3}  \partial_x T\big). 
\end{align}
\end{subequations}
These are the basic governing equations for the core region. Using Eq.~\eqref{heat} we deduce an estimate
\begin{equation}
\rho\,\frac{T}{t} \sim \frac{T^{3/2}}{x^2}\, \rho^{1/3},
\end{equation}
which is combined with $\rho\sim \sqrt{\xi}\sim |x|^{1/2}/t^{1/3}$ to give an estimate, 
\begin{equation}
\label{inside}
T \sim |x|^\frac{14}{3}\,t^{-\frac{22}{9}},
\end{equation}
of the temperature in the hot core where thermal conductivity plays a significant role.  The estimates Eqs.~\eqref{outside} and \eqref{inside} are comparable at $|x|=X$ determined by
\begin{equation*}
X^\frac{14}{3}\,t^{-\frac{22}{9}} \sim X^{-\frac{1}{2}}\,t^{-\frac{1}{3}},
\end{equation*}
from which we determine the growth law, 
\begin{equation}
\label{size}
X\sim t^\frac{38}{93},
\end{equation}
for the size $X$ of the hot core. The scaled size decays as $\Xi \sim X\,t^{-\frac{2}{3}} \sim t^{-\frac{8}{31}}$. 

\begin{figure*}
	\begin{center}
		\includegraphics[width=5.5cm,angle=0]{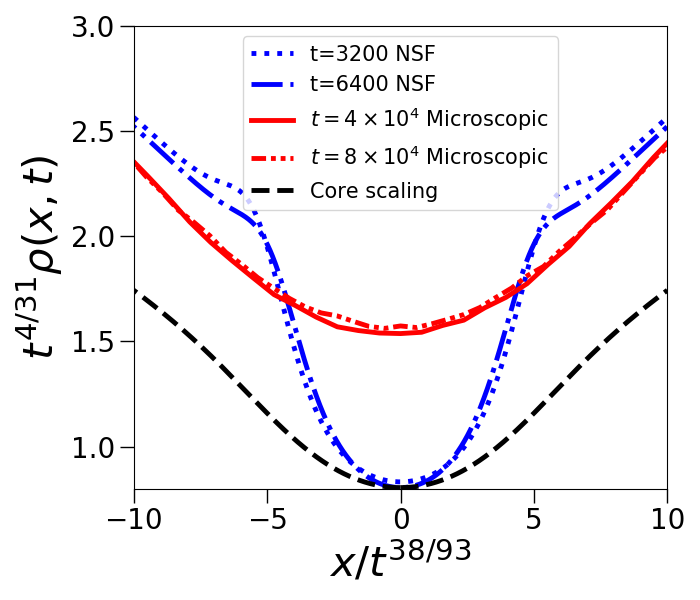}
		\includegraphics[width=5.5cm,angle=0]{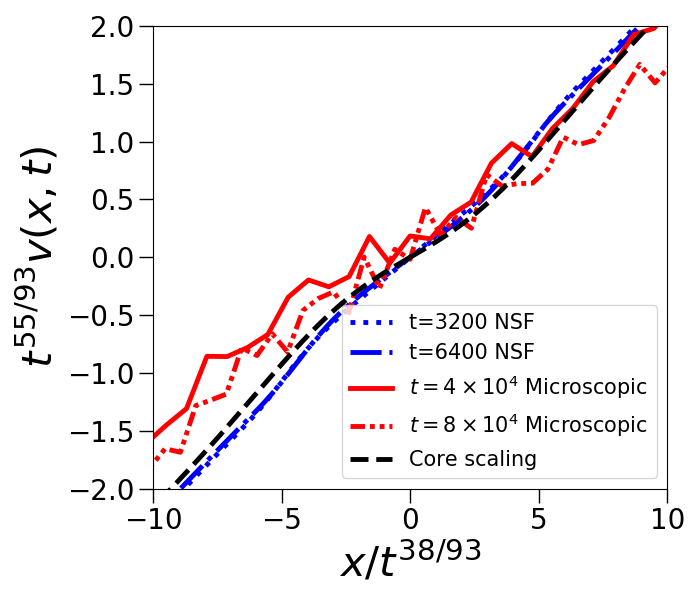}
		\includegraphics[width=5.5cm,angle=0]{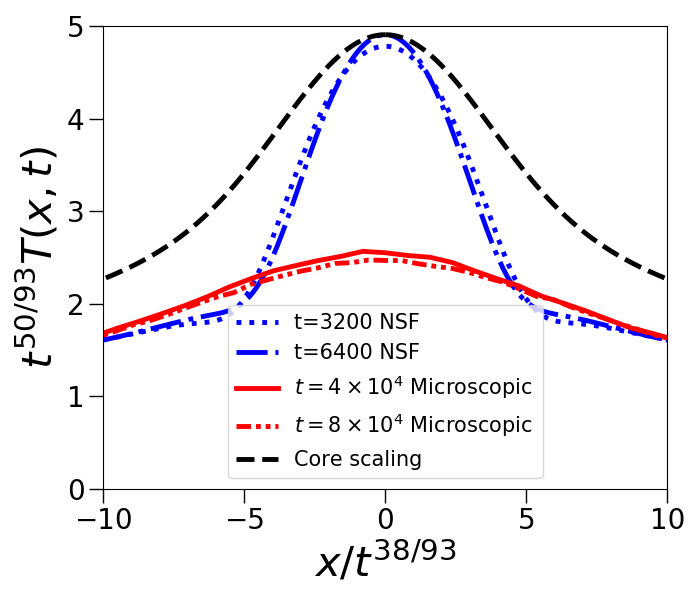}
		\caption{{Numerical verification of the core scaling forms Eqs.~(\ref{scaling:core})   for data for the fields $\rho,~v$ and  $T$ from both the microscopic simulations and NSF hydrodynamic numerics. The black dashed lines show the analytic scaling functions obtained from a numerical solution of Eqs.~(\ref{FV-eq}, \ref{F-V}).}}
		\label{Fig:coreprofile}
	\end{center}
\end{figure*}

 The temperature at the center of the explosion is estimated from $T_0\sim X^{-\frac{1}{2}}\,t^{-\frac{1}{3}}$ and Eq.~\eqref{size}. Similarly, the density at the center of the explosion is estimated from $\rho_0\sim X^{1/2}/t^{1/3}$ and Eq.~\eqref{size}. We thus arrive at the following 
asymptotic decay laws 
\begin{equation}
\label{T:origin}
T_0 \sim t^{-\frac{50}{93}}\,, \qquad \rho_0 \sim t^{-\frac{4}{31}}
\end{equation}
announced in the Introduction. 
In Fig.~\eqref{Fig:coreprofile} we check if  these decay forms  are satisfied  by the data from the NSF numerical solution as well as those from microscopic simulations. Especially for temperature it is clear that this form describes the data better than the TvNS form. Using Eq.~\eqref{v-core} we find that the velocity in the hot core scales as $X/t\sim t^{-\frac{55}{93}}$. 

The hydrodynamic variables in the hot core where dissipative effects play the dominant role exhibit scaling behaviors, but instead of $\xi=x/R$ we should use the scaled spatial coordinate $\eta=x/X$. Thus the hydrodynamic variables have a self-similar form
\begin{equation}
\label{scaling:core}
\rho = t^{-\frac{4}{31}} \widetilde{G}(\eta), \quad v = t^{-\frac{55}{93}} \widetilde{V}(\eta), \quad T = t^{-\frac{50}{93}} \widetilde{Z}(\eta)
\end{equation}
in the scaling region
\begin{equation}
\label{scaling:eta}
x\to \infty, \quad  t\to \infty, \quad \eta \sim x\,t^{-\frac{38}{93}} \to\text{finite}.
\end{equation}
In Fig.~\eqref{Fig:coreprofile} we verify that both the data from the NSF numerics and the microscopic simulations satisfy these scaling forms and give a better collapse of data than the TvNS scaling. 

We now combine the scaling ansatz Eqs.~\eqref{scaling:core}--\eqref{scaling:eta} with the governing equations \eqref{cont-eq-1}, \eqref{NS-zeta}, \eqref{heat}, setting for now $\mu=1.5,\ D_2=1$. Equation \eqref{NS-zeta} yields the most simple outcome, namely it becomes $(\widetilde{G}\widetilde{Z})'=0$ in the leading order. (Hereinafter prime denotes the differentiating with respect to $\eta$.) Thus $\widetilde{G}=1/\widetilde{Z}$, where for the moment we have set the integration constant to one (we discuss this point again later). Eq.~\eqref{cont-eq-1} then simplifies to 
\begin{subequations}
\begin{equation}
\label{GV-eq}
\widetilde{V}' - \tfrac{4}{31} =  \big(\widetilde{V}-\tfrac{38}{93}\eta\big) (\ln \widetilde{Z})',
\end{equation}
while Eq.~\eqref{heat} becomes 
\begin{equation}
\label{FV-eq}
\big(\widetilde{Z}^\frac{1}{6}\,\widetilde{Z}'\big)' + \tfrac{13}{93} =  \tfrac{3}{2}\big(\widetilde{V}-\tfrac{38}{93}\eta\big) (\ln \widetilde{Z})'.
\end{equation}
\end{subequations}
Comparing Eqs.~\eqref{GV-eq} and \eqref{FV-eq} we obtain 
\begin{equation*}
\big(\widetilde{Z}^\frac{1}{6}\,\widetilde{Z}'\big)' + \tfrac{13}{93} =  \tfrac{3}{2}\big(\widetilde{V}' - \tfrac{4}{31}\big),
\end{equation*}
which is integrated to give
\begin{equation}
\label{F-V}
\widetilde{V} = \tfrac{2}{9}\eta + \tfrac{2}{3} \widetilde{Z}^\frac{1}{6}\,\widetilde{Z}'.
\end{equation}
The integration constant vanishes since $\widetilde{V}(0)=0$ and $\widetilde{Z}'(0)=0$ due to symmetry. Inserting Eq.~\eqref{F-V} into Eq.~\eqref{FV-eq} we obtain
\begin{equation}
\label{F-eq}
\big(\widetilde{Z}^\frac{1}{6}\,\widetilde{Z}'\big)' + \tfrac{13}{93} =  \big(-\tfrac{26}{93}\eta+\widetilde{Z}^\frac{1}{6}\,\widetilde{Z}'\big) (\ln \widetilde{Z})'.
\end{equation}
Solving the above equation numerically [or equivlaently Eqs.~(\ref{FV-eq}, \ref{F-V})] with $\widetilde{Z}(0)$ fixed from the NSF numerical data at $t=6400$ and $\widetilde{Z}'(0)=0$ we find the scaling functions $\widetilde{G},\widetilde{V},\widetilde{Z}$. In Fig.~\eqref{Fig:coreprofile} we plot these and compare with the  data from NSF. 
We find qualitative agreement in the form of the scaling functions but not complete agreement. 
See Appendix~\eqref{app:core} for a more detailed derivation of the  equations for the scaling functions $\widetilde{G},\widetilde{V},\widetilde{Z}$, where we also introduce an additional scale factor $\lambda$ to make $\eta$ dimensionless. This gives us another fitting parameter  but does not lead to a significant improvement in our comparison with data.

\begin{figure}
	\begin{center}
\includegraphics[width=5.9cm,angle=0]{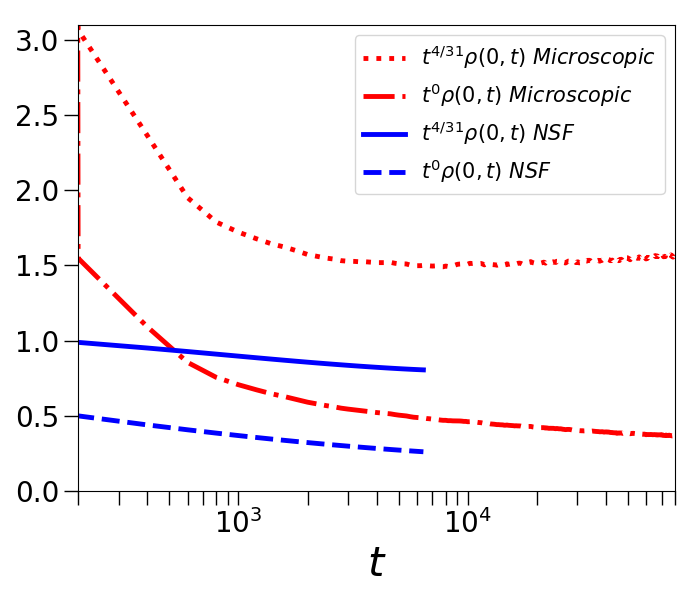}
\includegraphics[width=5.9cm,angle=0]{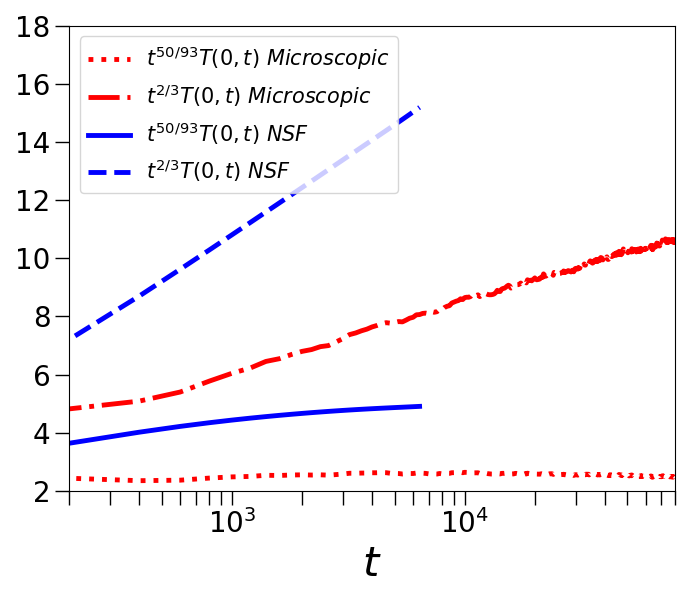}
\caption{{Plots showing evolution with time of the density and temperature at the centre of the blast for data from microscopic simulations and from NSF numerics. The scaling prefactors correspond to the predictions from TvNS and the core-scaling analysyis. We observe that for density it is difficult to differentiate between the two scaling forms while for temperature, the core-scaling prediction $T_0 \sim t^{-50/93}$ better describes the data than the TvNS prediction $T_0 \sim t^{-2/3}$.}} 
\label{Fig:centreprofile}
 	\end{center}
\end{figure}

The asymptotic behaviors of the velocity field is given by
\begin{subequations}
\begin{align}
\label{V-origin}
&\widetilde{V} \simeq \tfrac{4}{31} \eta \quad\qquad \qquad\text{when} \quad |\eta|\ll 1, \\
\label{V-far}
&\tfrac{2}{9}\eta - \widetilde{V} \sim \eta^{-\frac{19}{12}} \qquad\text{when}\quad \eta\gg 1,
\end{align}
\end{subequations}
where the form Eq.~\eqref{V-origin} follows from Eq.~\eqref{GV-eq}. The slope predicted by Eq.~\eqref{V-origin} is a bit smaller than the slope $\frac{2}{9}$ in the blast region.  The asymptotic Eq.~\eqref{V-far} follows from Eq.~\eqref{F-V}  and 
\begin{equation}
\label{F-far}
\widetilde{Z}\sim \eta^{-1/2}   \qquad\text{when}\quad \eta\gg 1
\end{equation}
This asymptotic can be derived from Eq.~\eqref{F-eq}. Indeed, $\widetilde{Z}^\frac{1}{6}\,\widetilde{Z}'$ is asymptotically negligible, and hence Eq.~\eqref{F-eq} simplifies to $(\ln \widetilde{Z})'=(2\eta)^{-1}$ leading to Eq.~\eqref{F-far}.

{\bf Matching of the inner and outer solutions}: The asymptotic behavior Eq.~\eqref{F-far} yields $\widetilde{G}=\widetilde{Z}^{-1}\sim\eta^{1/2}$ matching the small $\xi$ behavior Eq.~\eqref{G-0}. For the temperature in the outer region we have [ignoring numerical factors]
\begin{align}
T \sim \left(\frac{x}{t}\right)^2 Z(\xi) \sim  \left(\frac{x}{t}\right)^2  \left(\frac{x}{t^{2/3}}\right)^{-\frac{5}{2}} \sim x^{-1/2} t^{-1/3}, 
\end{align}
while in the inner region
\begin{align}
T = \f{\widetilde{Z}(\eta)}{t^{50/93}}  \sim t^{-50/93} \left(\f{x}{t^{38/93}}\right)^{-1/2}  
    \sim x^{-1/2} t^{-1/3}.
\end{align}
Thus we have perfect matching in the overlap of the inner and outer regions.
Similarly for the density in the outer region
\begin{align}
\rho = {G}(\xi)\sim (x/t^{2/3})^{1/2} \sim x^{1/2} t^{-1/3},
\end{align}
while in the inner region 
\begin{align}
\rho = t^{-4/31} \widetilde{G}(\eta) \sim  t^{-4/31} (x/t^{38/93})^{1/2} \sim
        x^{1/2} t^{-1/3},
\end{align}
where we used $\widetilde{G}=1/\widetilde{Z}$. Again we get perfect matching.

\section{Conclusions}
\label{sec:conclusion}
The present work examines the question of the connection between the atomistic description of matter with particles obeying Newton's equations and the continuum hydrodynamic description due to Euler, Navier, Stokes and Fourier. The evolution of a blast wave initial condition in a one-dimensional gas with ideal gas equation of state was studied. The system can be described by the fields $\rho(x,t), v(x,t), T(x,t)$ of mass density, flow velocity and temperature respectively, which correspond to the  conservation  of mass, momentum and energy in the system. The so-called TvNS solution of the hydrodynamic Euler equations predicts  self-similar scaling form for the fields at long times. We compared these predictions with what one observes in direct molecular dynamics simulations of a gas composed of hard point particles with alternating masses. 

Our main results can be summarized as follows:
\begin{enumerate}
\item The equal mass hard particle gas and the AHP gas have the same equilibrium properties. However their dynamical properties are completetely different. For the blast wave initial conditions, the equal mass gas shows a ballistic evolution 
with a scaling $x \sim t$, while the non-integrability of the AHP gas leads to the sub-ballistic scaling with $ x \sim t^{2/3}$. 
\item We obtained the exact TvNS scaling solution  of the 1D Euler equation for the ideal gas. The position of the shock is given by $R(t)=[1071 E t^2/(152 \rho_\infty)]^{2/3}$ and depends only on the total energy, $E$, of the blast and the ambient density, $\rho_\infty$. The scaling functions for the three hydrodynamic fields are found explcitly in terms of the scaling variable $\xi=x/R(t)$. 
\item For the AHP gas, we find a remarkably close agreement between the microscopic simulation results and the TvNS predictions.  This includes a matching of the location of the front $R(t)$ and of the scaling functions over most of the range of the scaling variable $\xi$, except for a core region near the origin whose size shrinks (when measured in $\xi$) at larger times. 
\item  We  obtained results both for the ensemble averaged fields and the empirical fields and find agreement between them. We also showed that it is important to use a microcanonical ensemble --- otherwise the energy fluctuations in a canonical ensemble lead to a broadening of the shock front.
\item We pointed out that the deviations from the TvNS scaling at the core occurs due to the contribution of  heat conduction that becomes non-negligible in this region. A  numerical and analytical study of the Navier-Stokes-Fourier equations provided us a detailed understanding of the distinct scaling form in the core region, whose size scales as $X\sim t^{38/93}$.  We analytically obtained the scaling forms for the three fields in the core region and verified these numerically. The divergence of the temperature field at the origin and the vanishing of the density field, which are  unphysical predictions of the TvNS solution,  are avoided in the solutions of the  dissipative equations.
\end{enumerate}

We note that earlier studies~\cite{Barbier2015,Barbier2015a,Barbier2016,Joy2017,Joy2018,Joy2019}  on hard sphere fluids in two and three dimensions were unable to obtain the clear and convincing agreeement between simulations and the TvNS solution (even in the non-core region) that our work has obtained.  Some possible reasons for this are: (i) for our system, an ideal gas, the equilibrium properties including the equation of state are known exactly and we are able to obtain an exact solution of the TvNS equations; (ii) simulations in the 1D AHP gas can be done very accurately, the system has good ergodicity properties, and large system sizes, long times can be easily studied.  

The complete understanding of the shrinking (under TvNS scaling) core region remains an open problem. A complete agreement between microscopics and hydrodynamic theory becomes a somewhat difficult question because of our lack of knowledge of the dissipation terms. In particular, the well known phenomenon of anomalous heat transport  in one-dimensional systems (with mass, momentum and energy conserved)  means that the thermal conductivity $\kappa$ in the hydrodynamic equation is not given by the standard Green-Kubo formula (for a recent resolution see \cite{saito2020}).  In conclusion, the present study finds a striking agreement (though not yet complete) between the observed evolution of profiles of conserved fields following a blast, obtained from the microscopic dynamics of an interacting one-dimensional fluid and that obtained from hydrodynamics.

\section{Acknowledgments}
We thank Anupam Kundu, R. Rajesh, Sriram Ramaswamy, Samriddhi Sankar Ray and Vishal Vasan for useful discussions.
We acknowledge support of the Department of Atomic Energy, Government of India, under project no.12-R$\&$D-TFR-5.10-1100.  The numerical simulations were done on the \textit{Contra} and \textit{Tetris} clusters of ICTS-TIFR

\bibliographystyle{unsrt}
\bibliography{references}

\begin{thebibliography}{10}

\bibitem{LandauBook}
L.~D. Landau and E.~M. Lifshitz.
\newblock {\em Fluid Mechanics}.
\newblock Pergamon Press, New York, 1987.

\bibitem{ZeldovichBook}
Y.~B. Zel'dovich and Y.~P. Raizer.
\newblock {\em Physics of Shock Waves and High-Temperature Hydrodynamic
  Phenomena}.
\newblock Academic Press, New York, 1967.

\bibitem{Taylor19501}
G.I. Taylor.
\newblock The formation of a blast wave by a very intense explosion {I}.
  {T}heoretical discussion.
\newblock {\em Proceedings of the Royal Society of London. Series A.
  Mathematical and Physical Sciences}, 201(1065):159--174, 1950.

\bibitem{Taylor19502}
G.I. Taylor.
\newblock The formation of a blast wave by a very intense explosion. - {II}.
  {T}he atomic explosion of 1945.
\newblock {\em Proceedings of the Royal Society of London. Series A.
  Mathematical and Physical Sciences}, 201(1065):175--186, 1950.

\bibitem{VonNeumann1963}
J.~Von~Neumann.
\newblock {\em John Von Neumann: Collected Works. Theory of games,
  astrophysics, hydrodynamics and meteorology}.
\newblock Number~6. Pergamon Press, 1963.

\bibitem{Sedov1946}
L.~I. Sedov.
\newblock Propagation of strong shock waves.
\newblock {\em Journal of Applied Mathematics and Mechanics}, 10:241--250,
  1946.

\bibitem{Sedov2014}
L.~I. Sedov.
\newblock {\em Similarity and Dimensional Methods in Mechanics}.
\newblock Elsevier Science, 2014.

\bibitem{BarenblattBook}
G.~I. Barenblatt.
\newblock {\em Scaling, Self-similarity, and Intermediate Asymptotics:
  Dimensional Analysis and Intermediate Asymptotics}.
\newblock Cambridge Texts in Applied Mathematics. Cambridge University Press,
  1996.

\bibitem{McKee1988}
J.~P. Ostriker and C.~F. McKee.
\newblock Astrophysical blastwaves.
\newblock {\em Rev. Mod. Phys.}, 60:1--68, Jan 1988.

\bibitem{Ditmire2001}
M.~J. Edwards, A.~J. MacKinnon, J.~Zweiback, K.~Shigemori, D.~Ryutov, A.~M.
  Rubenchik, K.~A. Keilty, E.~Liang, B.~A. Remington, and T.~Ditmire.
\newblock Investigation of ultrafast laser-driven radiative blast waves.
\newblock {\em Phys. Rev. Lett.}, 87:085004, Aug 2001.

\bibitem{Smith2005}
A.~S. Moore, D.~R. Symes, and R.~A. Smith.
\newblock Tailored blast wave formation: Developing experiments pertinent to
  laboratory astrophysics.
\newblock {\em Physics of Plasmas}, 12(5):052707, 2005.

\bibitem{Kellay2009}
J.~F. Boudet, J.~Cassagne, and H.~Kellay.
\newblock Blast shocks in quasi-two-dimensional supersonic granular flows.
\newblock {\em Phys. Rev. Lett.}, 103:224501, Nov 2009.

\bibitem{Kellay2013}
J.~F. Boudet and H.~Kellay.
\newblock Unstable blast shocks in dilute granular flows.
\newblock {\em Phys. Rev. E}, 87:052202, May 2013.

\bibitem{Antal2008}
T.~Antal, P.~L. Krapivsky, and S.~Redner.
\newblock Exciting hard spheres.
\newblock {\em Phys. Rev. E}, 78:030301, Sep 2008.

\bibitem{Jabeen2010}
Z~Jabeen, R.~Rajesh, and P.~Ray.
\newblock Universal scaling dynamics in a perturbed granular gas.
\newblock {\em {EPL} (Europhysics Letters)}, 89(3):34001, Feb 2010.

\bibitem{Barbier2015}
M.~Barbier, D.~Villamaina, and E.~Trizac.
\newblock Blast dynamics in a dissipative gas.
\newblock {\em Phys. Rev. Lett.}, 115:214301, Nov 2015.

\bibitem{Barbier2015a}
M.~Barbier.
\newblock Kinetics of blast waves in one-dimensional conservative and
  dissipative gases.
\newblock {\em Journal of Statistical Mechanics: Theory and Experiment},
  2015(11):P11019, nov 2015.

\bibitem{Barbier2016}
M.~Barbier, D.~Villamaina, and E.~Trizac.
\newblock Microscopic origin of self-similarity in granular blast waves.
\newblock {\em Physics of Fluids}, 28(8):083302, 2016.

\bibitem{Joy2017}
J.~P. Joy, S.~N. Pathak, D.~Das, and R.~Rajesh.
\newblock Shock propagation in locally driven granular systems.
\newblock {\em Phys. Rev. E}, 96:032908, Sep 2017.

\bibitem{Joy2018}
J.~P. Joy, S.~N. Pathak, and R.~Rajesh.
\newblock Shock propagation following an intense explosion: comparison between
  hydrodynamics and simulations.
\newblock 2018.

\bibitem{Joy2019}
J.~P. Joy and R.~Rajesh.
\newblock Shock propagation in the hard sphere gas in two dimensions:
  comparison between simulations and hydrodynamics, 2019.

\bibitem{Garrido2001}
P.~L. Garrido, P.~I. Hurtado, and B.~Nadrowski.
\newblock Simple one-dimensional model of heat conduction which obeys
  fourier’s law.
\newblock {\em Phys. Rev. Lett.}, 86(24):5486, 2001.

\bibitem{Dhar2001}
A~Dhar.
\newblock Heat conduction in a one-dimensional gas of elastically colliding
  particles of unequal masses.
\newblock {\em Phys. Rev. Lett.}, 86:3554--3557, Apr 2001.

\bibitem{Grassberger2002}
P~Grassberger, W~Nadler, and L~Yang.
\newblock Heat conduction and entropy production in a one-dimensional
  hard-particle gas.
\newblock {\em Phys. Rev. Lett.}, 89:180601, Oct 2002.

\bibitem{Casati2003}
G.~Casati and T.~Prosen.
\newblock Anomalous heat conduction in a one-dimensional ideal gas.
\newblock {\em Physical Review E}, 67(1):015203, 2003.

\bibitem{Cipriani2005}
P~Cipriani, S~Denisov, and A~Politi.
\newblock From anomalous energy diffusion to levy walks and heat conductivity
  in one-dimensional systems.
\newblock {\em Physical review letters}, 94(24):244301, 2005.

\bibitem{Chen2014}
S.~Chen, J.~Wang, G.~Casati, and G.~Benenti.
\newblock Nonintegrability and the {F}ourier heat conduction law.
\newblock {\em Physical Review E}, 90(3):032134, 2014.

\bibitem{Hurtado2016}
P.~I. Hurtado and P.~L. Garrido.
\newblock A violation of universality in anomalous {F}ourier’s law.
\newblock {\em Scientific reports}, 6(1):1--10, 2016.

\bibitem{lepri2020}
S.~Lepri, R.~Livi, and A.~Politi.
\newblock Too close to integrable: Crossover from normal to anomalous heat
  diffusion.
\newblock {\em Phys. Rev. Lett.}, 125:040604, Jul 2020.

\bibitem{Narayan2002}
O.~Narayan and S.~Ramaswamy.
\newblock Anomalous heat conduction in one-dimensional momentum-conserving
  systems.
\newblock {\em Physical review letters}, 89(20):200601, 2002.

\bibitem{Van2012}
H.~Van~Beijeren.
\newblock Exact results for anomalous transport in one-dimensional hamiltonian
  systems.
\newblock {\em Physical review letters}, 108(18):180601, 2012.

\bibitem{Mendl2013}
C.~B. Mendl and H.~Spohn.
\newblock Dynamic correlators of {F}ermi-{P}asta-{U}lam chains and nonlinear
  fluctuating hydrodynamics.
\newblock {\em Physical review letters}, 111(23):230601, 2013.

\bibitem{Spohn2014}
H~Spohn.
\newblock Nonlinear fluctuating hydrodynamics for anharmonic chains.
\newblock {\em Journal of Statistical Physics}, 154:1191--1227, 2014.

\bibitem{Mendl2014}
C.~B. Mendl and H.~Spohn.
\newblock Equilibrium time-correlation functions for one-dimensional hard-point
  systems.
\newblock {\em Physical Review E}, 90(1):012147, 2014.

\bibitem{Das2014}
S.~G. Das, A.~Dhar, K.~Saito, C.~B. Mendl, and H.~Spohn.
\newblock Numerical test of hydrodynamic fluctuation theory in the
  {F}ermi-{P}asta-{U}lam chain.
\newblock {\em Physical Review E}, 90(1):012124, 2014.

\bibitem{Mendl2017}
C.~B. Mendl and H.~Spohn.
\newblock Shocks, rarefaction waves, and current fluctuations for anharmonic
  chains.
\newblock {\em Journal of Statistical Physics}, 166(3-4):841–875, Oct 2016.

\bibitem{RiemannBook}
E.~F. Toro.
\newblock {\em Riemann Solvers and Numerical Methods for Fluid Dynamics: A
  Practical Introduction}.
\newblock Springer, Berlin, 2009.

\bibitem{Kadanoff1995}
Y.~Du, H.~Li, and L.~P. Kadanoff.
\newblock Breakdown of hydrodynamics in a one-dimensional system of inelastic
  particles.
\newblock {\em Phys. Rev. Lett.}, 74:1268--1271, Feb 1995.

\bibitem{Hurtado2006}
P.~I. Hurtado.
\newblock Breakdown of hydrodynamics in a simple one-dimensional fluid.
\newblock {\em Physical review letters}, 96(1):010601, 2006.

\bibitem{krapivskybook}
P.~L. Krapivsky, S.~Redner, and E.~Ben-Naim.
\newblock {\em A kinetic view of statistical physics}.
\newblock Cambridge University Press, 2010.

\bibitem{maccormack1982}
R.~W. MacCormack.
\newblock A numerical method for solving the equations of compressible viscous
  flow.
\newblock {\em AIAA journal}, 20(9):1275--1281, 1982.

\bibitem{saito2020}
K.~Saito, M.~Hongo, A.~Dhar, and S.~Sasa.
\newblock Microscopic theory of the fluctuating hydrodynamics in nonlinear
  lattices.
\newblock {\em arXiv:2006.15570}, 2020.

\end{thebibliography}

\begin{appendix}

\section{Derivation of radial Euler equations and the Rankine-Hugoniot conditions}
\label{app:idealgas}
We start with the Euler equations corresponding to conservation of mass, momentum and energy: 
\begin{align}
&\p_t \rho +\nabla.(\rho {\bf{v}}) =0,  \label{e1} \\
&\rho (\p_t  v + {\bf v}.\nabla {\bf v}) +  \nabla P  =0, \label{e2} \\
& \rho (\p_t e + {\bf v}.\nabla   e ) +  \nabla . (P {\bf v})  =0. \label{e3}
\end{align}
where  $e=v^2/2+\epsilon$ and $\epsilon$ is the internal energy per unit mass of the gas. From the equilibrium thermodynamics of the gas we are provided a relation $P=P(\rho,\epsilon)$ which then gives us a complete set of equations for the independent fields, say $\rho,{\bf v}$ and $P$. Using the equation of state $P=P(\rho,T)$ we can also replace the pressure field by the temperature field. We denote the entropy per unit mass as $s=s(\rho,\epsilon)$. 

Assuming radial symmetry of the solutions the fields depend only on the the radial coordinate $r$ and we can write the  above equations in the form:
\begin{align}
&D_t \rho + \rho \partial_r  v +\frac{d-1}{r}\,\rho v  = 0\\
&D_t v +\frac{1}{\rho}\, \partial_r P = 0 \\
&D_t \epsilon - \f{P}{\rho} D_t \rho  = T D_t s =0,
\end{align}
where we have defined the fluid derivative $D_t=\p_t+v_r\p_r$ and the last equation follows on using $e=v^2/2+\epsilon$ along with the first two  conservation equations. So far, the results are quite general. For the case of the ideal gas we have $s=\ln (P/\rho^\gamma)$, up to a constant.  
This then gives us the equations Eq.~(\ref{cont-eq}, \ref{E-eq}, \ref{entropy-eq}) in the main text.

To obtain the so-called Rankine-Hugoniot boundary conditions at the shock we consider the time derivative of the conserved quantities inside a thin tube, along any radial direction, and that just spans across the shock (say from $R-\Delta$ to $R+\Delta$ with $\Delta \to 0$). Using the Leibniz rule for differentiation under the integral sign, and defining $\dot{R}=U$, we then get the equations: 
\begin{align}
&\f{\rho(R) v(R)}{\rho(R)-\rho_\infty }=U,  \\
&\f{ \rho(R) v^2(R) + P(R)}{\rho(R) v(R)}=U,  \\
&\f{ \rho(R) v(R) e(R) + P(R) v(R)}{\rho(R) e(R)}=U,
\end{align}
where $\rho(R),v(R),e(R)$ denote the values of the fields just behind the shock position. The density field ahead of the shock is taken as $\rho_\infty$, while the pressure and velocity are assumed to be negligible and set to zero. Using the definition $e=v^2/2+\epsilon$, where for the ideal gas one has $\epsilon=P/[\rho (\gamma-1)]$, we  see that the above equations gives us three equations for the fields $\rho(R),v(R),e(R)$ which can be solved in terms of $U$. These precisely give Eqs.~\eqref{RH}. 

Similarly, we can derive Eq.~\eqref{ZV} by considering the time derivative of the conserved energy inside a thin tube, along any radial direction, and that  spans from   $0$ to $r=\xi R(t)$. Again using the  Leibniz rule for differentiation under the integral sign we get
\begin{align}
\xi \dot{R} [\rho(R \xi) e(R \xi)] = [\rho(R\xi) e(R\xi) + P(R\xi)] v(R\xi).  \label{eq:Eeqderiv}
\end{align}
Using  the facts that $\dot{R}=\delta R(t)/t$, $e=v^2/2+ \epsilon,~\epsilon=T/[\mu (\gamma-1)],~P=\rho T/\mu$ and the scaling ansatz $v=\delta (r/t) V(\xi)$ and $T=(\mu \delta^2/\gamma) (r^2/t^2) Z(\xi)$,  Eq.~\eqref{eq:Eeqderiv} immediately leads to  Eq.~\eqref{ZV}.

\section{TvNS scaling of the hydrodynamic equations}
\label{sec:scaling}
Here we provide arguments, for the 1D fluid, that lead to the  TvNS scaling forms of the long time self-similar solutions of the hydrodynamic equations. For simplicity, we will consider only the Euler equations without the dissipation terms. One can easily show that the dissipation terms do not contribute in this scaling regime. However, as we discuss in the next section, the dissipative terms are non-negligible in the core of the blast and lead to a different scaling form.

 Let us assume the solutions to be of the form:
\begin{align}
\rho(x,t) &= \rho_\infty + t^b\widetilde{\rho}(xt^a),~ \label{eq:rhoscale}\\
v(x,t) &= t^c\widetilde{v}(xt^a),~\label{eq:uscale}\\ 
e(x,t) &= t^d\widetilde{e}(xt^a).~\label{eq:escale} 
\end{align}
Now substituting these scaling forms into Eqs.~(\ref{tNS1}-\ref{tNS3}) without the dissipative terms on the right hand sides, and defining the scaling variable $z=x t^a$, we arrive at the following equations:
\begin{align}
&b\widetilde{\rho} + a z \widetilde{\rho}^\prime + t^{a+c+1}(\widetilde{\rho}\widetilde{v})^\prime+\rho_\infty t^{a-b+c+1}\widetilde{v}^\prime= 0,~ \label{rhoscaletilde}\\
&(b+c)\widetilde{\rho}\widetilde{v} + a z (\widetilde{\rho}\widetilde{v})^\prime +\rho_\infty c t^{-b}\widetilde{v}+
\rho_\infty a z t^{-b}\widetilde{v}^\prime \nn \\ &~~~~~~~~+ 2 t^{a-c+d+1}( \widetilde{\rho}\widetilde{e})^\prime   
+2\rho_\infty t^{a+d-c-b+1}\widetilde{e}^\prime=0,~\label{uscaletilde}  \\
&(b+d)\widetilde{\rho}\widetilde{e} + a z ( \widetilde{\rho}\widetilde{e})^\prime +\rho_\infty dt^{-b}\widetilde{e}+ \rho_\infty a z t^{-b}\widetilde{e}^\prime \nn \\ & + 3t^{a+c+1}(\widetilde{\rho}\widetilde{e}\widetilde{v})^\prime
+ 3\rho_\infty t^{a+c-b+1}(\widetilde{e}\widetilde{v})^\prime  -t^{a+3c-d+1} (\widetilde{\rho}\widetilde{v}^3)^\prime \nn \\&~~~~~~~~~~~~~~~~~~~~~~~~~~~~~-\rho_\infty t^{a+3c-b-d+1}(\widetilde{v}^3)^\prime= 0.~\label{escaletilde} 
\end{align}
Requiring now that there exists  a scaling solution we set all powers of the time variable to zero, which then gives us the conditions $b=0$, $a+c+1=0$ and $d=2c$. We still need one condition to determine the exponents completely and, as  first noticed by Taylor~\cite{Taylor19501,Taylor19502}, the energy conservation condition turns out to be sufficient. 

The constancy of energy means that the integral
\begin{equation}
\int_{-\infty}^{\infty} dx \rho(x,t)e(x,t)  = E ,
\end{equation}
should be a constant, independent of time. Plugging in our scaling forms for $\rho$ and $e$, and making a change of variables,  the above gives 
\begin{equation}
t^{d-a}\int_{-\infty}^{\infty} dz \widetilde{\rho}(z) \widetilde{e}(z)  = E 
\end{equation}
Demanding time-independence then requires $d=a$ and together with the earlier equations we finally get:
\begin{equation}
a=-2/3,~b=0,c=-1/3,~ d=-2/3.~ \label{c4}
\end{equation}
This is in agreement to the TvNS prediction for one dimension and is consistent with the scalings that we have observed numerically for the hard particle system with alternating masses as well as the numerical solution of the hydrodynamic equations with dissipation. It is easy to show that with the above choice of the scaling solution, all the dissipation terms in the hydrodynamic equations scale as $t^{-1/3}$ and hence do not contribute to the solution in the long time limit. 

We note here that the above self-similar solution was obtained for initial conditions corresponding to a system with zero ambient temperature. For the case of a finite ambient temperature (and correspondingly a finite energy density $e_0$), a similar scaling analysis as that presented  in 
the previous paragraphs, leads to the scaling solution  $\rho(x,t)-\rho_\infty =\widetilde{\rho}(t^{-1}x), v(x,t)=t^{-1}\widetilde{v}(t^{-1}x), e(x,t)-e_0=t^{-1} \widetilde{e}(t^{-1}x)$. Another well-studied set-up is  where one starts with an initial condition with step profiles of all the conserved variables (the Riemann problem, see e.g \cite{Mendl2017}). In this case, one gets self-similar solutions of the form $\rho(x,t)=\widetilde{\rho}(t^{-1}x), v(x,t)=\widetilde{v}(t^{-1}x), e(x,t)=\widetilde{e}(t^{-1}x)$.

\section{Scaling solution in the core of the blast}
\label{app:core}
We consider the heat equation
\begin{equation}
\label{heatapp}
\rho\partial_t (T/2 \mu) = \partial_x (\kappa  \partial_x T) 
\end{equation}
with $\kappa = D_2\rho^{1/3} T^{1/2}$.  Near the origin, the ratio of the RHS to the LHS can be estimated from the TvNS solution as
\begin{align}
\approx \f{T^{1/2} t}{\rho^{2/3} x^2} \sim \f{1}{\xi^{5/4} \xi^{1/3} x} \sim \f{t^{19/18}}{x^{31/12}}.  
\end{align}
We see that this becomes large in the region $x \lesssim t^{38/93}$. Hence we expect a different scaling solution in the region $ x < t^{-a}$ where $a=-38/93$. 

We have already established $a=-38/93$ and $d=-50/93$ (see main text). Now we write the three hydrodynamic equations in the form:
\begin{align}
&\partial_t \rho + \partial_x (\rho v) = 0 \\
&\rho (\p_t + v \p_x) v + \p_x P = \p_x (\eta \p_x v) \\
& \rho (\p_t + v \p_x) \epsilon+ P\p_x v = \p_x (\eta v \p_x v)+ \p_x  (\kappa\partial_x T). 
\end{align}
For the ideal gas we use  $\epsilon = T/(2 \mu)$ and $P=\rho T/\mu$. Plugging in the scaling forms we then get: 
\begin{align}
	&t^{b-1} (b \widetilde{G} + a \eta \widetilde{G}') +  t^{a+b+c} \lambda (\widetilde{G} \widetilde{V})' = 0. \label{eq:sc1} \\
	&t^{b+c-1} ( c \widetilde{G} \widetilde{V} + a \eta \widetilde{G} \widetilde{V}') + t^{b+2c +a} \lambda \widetilde{G} \widetilde{V} \widetilde{V}'+ t^{a+b+d} \lambda \f{(\widetilde{G}\widetilde{Z})'}{\mu} \nn  \\
	& = t^{d/2+c+2 a} \lambda^2 (\widetilde{Z}^{1/2} \widetilde{V}')' \label{eq:sc2}\\
	& t^{b+d-1} \left[\f{d \widetilde{G}\widetilde{Z}}{2 \mu}+ \f{a \eta \widetilde{G} \widetilde{Z}'}{2 \mu}\right] + t^{b+c+d+a} \lambda\f{\widetilde{G} \widetilde{V} \widetilde{Z}'}{2 \mu} \nn \\ 
	&+ t^{a+b+c+d} \lambda \f{\widetilde{G} \widetilde{Z} \widetilde{V}'}{\mu} = t^{d/2+2c+2a} \lambda^2 (\widetilde{Z}^{1/2}\widetilde{V} \widetilde{V}')' \nn \\ &+  t^{b/3+3d/2+2 a} \lambda^2 (\widetilde{G}^{1/3} \widetilde{Z}^{1/2}  \widetilde{Z}')'. \label{eq:sc3}
	\end{align}
From Eq.~\eqref{eq:sc1} we get $c=-1-a=-55/93$ and then from Eq.~\eqref{eq:sc3} 
we get $b=3d/4+3a+3/2=-4/31$. The leading term in Eq.~\eqref{eq:sc2} then gives 
$(\widetilde{G}\widetilde{Z})'=0$ implying $\widetilde{G}=k/\widetilde{Z}$, where $k=\widetilde{G}(0)\widetilde{Z}(0)$ is a constant. Using this then leads to the following equations for $\widetilde{V}$ and $\widetilde{Z}$:
\begin{align}
&  \widetilde{V}'-\f{4}{31} + \left(\f{38\eta  }{93} - \widetilde{V} \right) \f{\widetilde{Z}'}{\widetilde{Z}} = 0, \label{eq:scp1} \\
&  \widetilde{V}'-\f{25}{93}- \f{1}{2} \left( \f{38 \eta}{93} -   \widetilde{V} \right) \f{ \widetilde{Z}'}{\widetilde{Z}}  
  =  p \lambda^2 (\widetilde{Z}^{1/6}  \widetilde{Z}')', \label{eq:scp3}
\end{align}
where $p = \mu/k^{2/3}$. From these two equations we get
\begin{align}
\f{3 }{2} \widetilde{V}'-1/3=p \lambda^2 (\widetilde{Z}^{1/6}  \widetilde{Z}')'.
\end{align}
This can be integrated to give
\begin{align}
 \widetilde{V}= 2\eta/9+(2p\lambda^2 /3) \widetilde{Z}^{1/6}  \widetilde{Z}'.
\end{align}
Substituting this back in Eq.~\eqref{eq:scp1} we get an equation for $\widetilde{Z}$:
\begin{align}
\lambda^2 p (\widetilde{Z}^{1/6}  \widetilde{Z}')'+ \left(\f{26 \eta}{93} - \lambda^2 p \widetilde{Z}^{1/6}  \widetilde{Z}' \right) \f{\widetilde{Z}'}{\widetilde{Z}}+\f{13}{93} = 0.
\end{align}
The constants $\widetilde{G}(0),\widetilde{V}(0),\widetilde{Z}(0)$ and $\lambda$ should be fixed such that the fileds at large $\eta$ match with the small $\xi$ TvNS solution.

\end{appendix}

\end{document}